\documentclass{aastex631}

\usepackage{subfigure}
\usepackage{color}

\usepackage{ulem}

\begin{document}

\title{
Circumbinary disk spectra irradiated by two central accretion disks in a binary black hole system
%
}


\correspondingauthor{Kimitake Hayasaki}
\email{yunewoo.lee@chungbuk.ac.kr, kimi@chungbuk.ac.kr}

\author{Yunewoo Lee}
\affiliation{Department of Astronomy and Space Science, Chungbuk National University, Republic of Korea}

\author{Atsuo T. Okazaki}
\affiliation{Hokkai-Gakuen University, Toyohira-ku, Sapporo 062-8605 Japan}


\author{Kimitake Hayasaki}
\affiliation{Department of Astronomy and Space Science, Chungbuk National University, Republic of Korea}
\affiliation{Department of Physical Sciences, Aoyama Gakuin University, Sagamihara 252-5258, Japan}

\begin{abstract}
We study the effect of irradiation from two accretion disks (minidisks) around respective black holes of stellar to intermediate masses in a circular binary on the spectrum of a circumbinary disk (CBD) surrounding them. We assume the CBD to be a standard disk and adopt the orbit-averaged irradiation flux because the viscous timescale is much longer than the orbital period. We then solve the energy equation both analytically and numerically to compute the CBD temperature distribution and the corresponding disk spectrum. 
We find that the analytically calculated spectra are in good agreement with the numerical ones. The CBD spectrum is almost independent of the binary mass ratio. We also find that the combined spectra of two minidisks and the CBD have double peaks, one peak in the soft X-ray band and the other in the infrared (IR) band. The former peak comes from the two minidisks, while the latter peak from the CBD. The observed flux density increases with frequency as $\nu^{1/3}$ towards the soft X-ray peak, while it decreases with frequency away from the IR peak as $\nu^{-5/3}$. The latter feature is testable with near-IR observations with Subaru and JWST.
\end{abstract}


%
\section{Introduction} 
\label{sec:intro}
%


Binary black holes (BBHs) are composed of two black holes in a binary system. They are of particular interest in astrophysics because they provide key insights into black hole mergers and the growth from a seed black hole to a supermassive black hole \citep{1980Natur.287..307B}. The gravitational interactions within BBHs lead to the emission of gravitational waves, as predicted by general relativity (GR), and their detection has been a major breakthrough in observational astronomy \citep{2016PhRvL.116f1102A}. BBHs are typically formed by the stellar collapse of massive binary systems and by dynamical interactions in dense stellar environments such as globular clusters \citep{2002ApJ...576..894M}. Sub-solar-mass BBHs can be formed by three-body interactions between the primordial black holes
\citep{1998PhRvD..58f3003I,2016PhRvL.117f1101S}.


The entire binary system is surrounded by the circumbinary disk (CBD), which orbits the common center of mass of the two black holes. The tidal-resonant interactions between the CBD and the binary black hole lead to the formation of gaps or cavities within the CBD \citep{artymowicz_dynamics_1994,2014ApJ...783..134F,dorazio_transition_2016,2023ApJ...949L..30D}, and lead to the orbital decay of the BBH by angular momentum transfer \citep{2009PASJ...61...65H,2009MNRAS.393.1423C,2009ApJ...700.1952H,hayasaki_gravitational_2013}. In addition, the CBD material falls inward from the two points at the inner edge of the CBD, forming a circumprimary disk (CPD) around the more massive (primary) black hole and a circumsecondary disk (CSD) around the less massive (secondary) black hole, and eventually accretes onto the black holes through these minidisks (e.g., \citealt{hayasaki_supermassive_2008}).

The thermal emission from the CBD has been studied by \cite{rafikov_structure_2013,rafikov_generalized_2016} based on the long-term evolution of the CBD, and \cite{2014ApJ...785..115R} has studied the non-thermal X-ray emission due to the shock formed by the interaction between the CBD and two minidisks. The disk-binary interaction can also cause periodic accretion \citep{hayasaki_binary_2007,2008ApJ...672...83M,2009MNRAS.393.1423C,2012ApJ...755...51N,2013MNRAS.436.2997D,2015ApJ...807..131S}. Recently, GR effects have also been found to give the light curve variations of BBH systems: a special relativistic Doppler boost in the emission from each minidisk of a rapidly orbiting binary at relativistic speeds \citep{2015Natur.525..351D}, 
the quasi-periodic modulation of the structure of two minidisks in a coalescing black hole \citep{2018ApJ...853L..17B}, and the GR precession of the semi-major axis of the binary \citep{2024arXiv240507897D}. These signals can help to identify BBH candidates in electromagnetic surveys.



The study of X-ray self-irradiated disks has advanced our understanding of the accretion dynamics and radiative processes of binary systems. The pioneering work of \cite{1981PASJ...33..365H} explored the effects of X-ray irradiation on accretion disks and established a basic model for how X-ray photons from a central source of the disk can heat the outer regions of the disk, thereby influencing its emission properties. \cite{1992PASJ...44..663F} and \cite{1993PASJ...45..443S} extended this model to include the effects of self-irradiation, where X-rays emitted by the disk itself are reprocessed by the disk material, leading to thermal and emission profiles that differ from the standard disk spectrum (see also \citealt{kato_black-hole_1998} for a review).

Subsequent studies, such as that of \cite{1998MNRAS.293L..42K}, have provided detailed analyses of the light curves of soft X-ray transients, showing that self-irradiation can significantly alter the observational properties of these systems. Furthermore, \cite{matsumoto_irradiated_1998} studied the Galactic supersoft X-ray source RX J0019.8+2156 and showed that reprocessed radiation from the accretion disk and companion star could account for the observed fluxes in UV and optical wavelengths, highlighting the role of irradiation in shaping the spectral energy distributions of such systems. Recently, a long-term multi-wavelength study of the microquasar GRS 1915+105 revealed that the soft X-ray photons coming from the inner region of the disk are reprocessed to thermalize the outer part of the disk \citep{rahoui_long-term_2010}. It has also been argued that, in ultraluminous X-ray sources, the X-ray flux from the inner source is reprocessed in the outer regions of the accretion disk, dominating the spectrum at optical and UV wavelengths (e.g., \citealt{copperwheat_optical_2005,copperwheat_irradiation_2007,2014MNRAS.444.2415S})


While supermassive BBHs naturally form CBDs in the galactic nuclei, BBHs composed of stellar-mass black holes and intermediate-mass black holes (IMBHs) must exist in a gaseous environment to accompany the CBDs \citep{hayasaki_gravitational_2013,2016PASJ...68...66H}. If such BBHs were to plunge into a gas cloud, such as a molecular cloud, they would end up with a triple disk, consisting of two minidisks and a CBD surrounding them.  In this case, X-ray photons from the CPD and the CSD would irradiate the outer part of the CBD, modifying the emission properties of the CBD. Figure 1 shows a schematic representation of X-ray irradiation of the outer part of the CBD from near the inner edge of these two minidisks.  It is assumed that  the orbital plane of the binary and the CBD are aligned.

However, little is known about how X-rays from the inner region of the CPD and CSD theoretically affect the CBD surface in a binary black hole system. In this paper, we study analytically and numerically the effect of irradiation from the two minidisks on the CBD. In Section 2, we construct the basic model of the irradiated CBD. Note that the detailed descriptions are given in the Appendix. Section 3 provides the method to solve the energy equation of the CBD analytically and numerically. In Section 4, we present their solutions and the combined spectra of two minidisks and the CBD, called triple disk spectra, and describe how our models are tested by near-infrared (NIR) observations with Subaru and JWST. Sections 5 and 6 are devoted to discussion and conclusions, respectively.

\begin{figure*}[htbp]
    \centering    \includegraphics[scale=0.25]{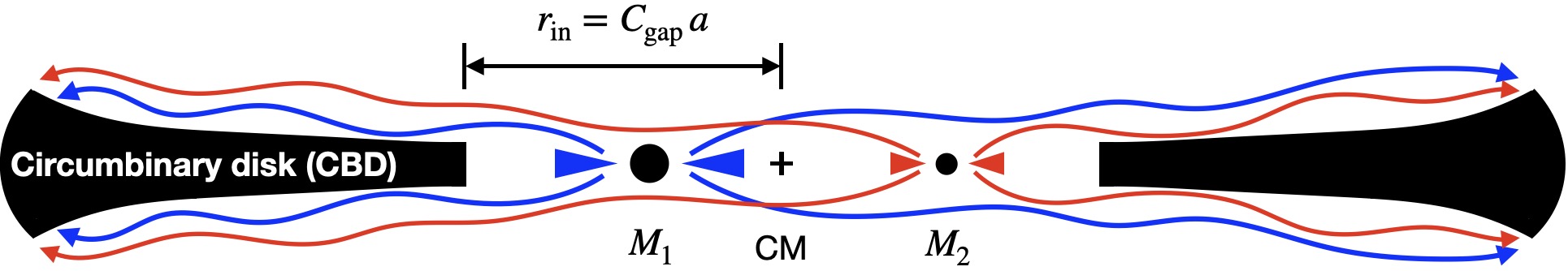}
    \caption{ 
    Edge-on schematic of a binary black hole system with two minidisks around the primary and secondary black holes and the CBD surrounding them. The primary and secondary black holes are indicated by black filled circles. The accretion disks around the primary and secondary black holes are CPD and CSD, respectively. The blue and red wavy lines represent photons emitted from the CPD and the CSD toward the surface of the CBD, respectively. The radial distance from the center of mass to the inner edge of the CBD is given by $r_{\rm in}=C_{\rm gap}a$, where $C_{\rm gap}$ is the parameter to measure the gap size between the CBD and the binary, and $a$ is the binary semimajor axis. 
    }
    \label{fig:tripledisk}
\end{figure*}
%
\section{Model} 
\label{sec:model}
%
We aim to investigate the effect of irradiation from the two minidisks, which is absorbed and re-emitted at the surface of the CBD, on the disk spectrum of the CBD. First, we assume that the two minidisks and the CBD are one-dimensional, axisymmetric, and steady-state standard disks.
The binary has a mass $M$  and a circular orbit with a semi-major axis $a = a_0\,r_{\rm S}$, where $a_0$ is the parameter to measure the semi-major axis. Also, 
\begin{eqnarray}
    r_{\rm S}
    &&
    =\frac{2GM}{c^2}
    \nonumber \\
    &&
    =3.0\times10^{7}\,{\rm cm}\left(\frac{M}{100\,M_\odot}\right).
    \label{eq:rs}
\end{eqnarray}
is the Schwarzschild radius of the black hole with $M$, where $G$ is the gravitational constant and $c$ is the speed of light.

We note that irradiation from near the inner edge of each minidisk is unlikely to affect the spectrum of the outer part because the radius where the irradiation heating overcomes the viscous heating is much larger than the outer radius of the disk. Also, the energies of the photons generated at the inner edge of the CBD are in the optical wavelength and are much lower than those of the photons from the inner edge of two minidisks. This is because the disk temperature of each minidisk is higher than the CBD temperature because of $T\propto{r}^{-3/4}$ for the standard disk model \citep{1981ARA&A..19..137P}. Therefore, we consider the effect of irradiation from the inner edge of the CBD on the CBD spectrum to be tiny, and thus we neglect the irradiation heating rate due to the CBD inner edge in the energy equation. In this work, as shown in Figure~\ref{fig:tripledisk}, we examine the effect of irradiating the surface of the CBD from the inner edges of the two minidisks. We then calculate the spectrum of the entire binary system, including the minidisks and the CBD.

Since the CBD is optically thick, the radiative flux from both sides of the disk surface is locally proportional to the fourth power of the disk temperature according to the Stefan-Boltzmann law, i.e. the radiative cooling rate is given by
\begin{eqnarray}
     Q_{\rm rad}=2\sigma T^4,
     \label{eq:qrad}
\end{eqnarray}
where $\sigma$ is the Stefan-Boltzmann constant. According to the standard disk theory, the heating rate due to the viscous heating of the disk is given by 
\begin{eqnarray}
    Q_{\rm vis}
    = 
    \frac{3GM\dot M}{4\pi r^3},
    \label{eq:qvis}
\end{eqnarray}
for $r\gg{r_{\rm ISCO}}$ in the CBD, where $r_{\rm ISCO}=6GM/c^2$ is the radius at the innermost stable circular orbit (ISCO). Here we take the mass accretion rate as 
\begin{eqnarray}
    \dot{M}=\dot{m}\frac{L_{\rm Edd}}{c^2},
    \label{eq:mdot}
\end{eqnarray}
where $L_{\rm Edd}=4\pi GMc/\kappa$ is the Eddington luminosity with the opacity of the gas $\kappa$ and $\dot{m}$ is the ratio of mass accretion rate to $L_{\rm Edd}/c^2$. We take $\dot{m}=1$ throughout the paper.

Since the CBD viscous timescale $\tau_{\rm vis}$ is much longer than the binary orbital period $P_{\rm orb}$ (see equations~\ref{eq:vistime}-\ref{eq:bop} for details), the binary system with triple disk composed of CBD, CPD, and CSD can be in a quasi-steady state , which thus allows to impose on the following relation on the accretion rate between CBD, CPD, and CSD: 
\begin{eqnarray}
 \dot{M}=\dot{M}_1+\dot{M}_2.
\end{eqnarray}
Also, we assume that the mass accretion rate ratio is equal to the binary mass ratio $q$, i.e. $\dot{M}_2/\dot{M}_1=q$, where $\dot{M}_1$ and $\dot{M}_2$ are the CPD and CSD accretion rates, respectively. These relation provides 
\begin{eqnarray}
\dot{M}_1= \frac{1}{1+q}\dot{M},\,\,
\dot{M}_2=\frac{q}{1+q}\dot{M}.
\end{eqnarray}

The energy equation $Q_{\rm vis} + Q_{\rm irr}=Q_{\rm rad}$ is given by
\begin{eqnarray}
    \frac{3GM\dot M}{4\pi r^3}
+
&& 
    \frac{A_1L_1}{2\pi r} 
     \Biggr[
     \frac{d}{dr}\left(\frac{H}{r}\right)
     -
    \beta_1\left(\frac{r_{\rm in}}{r}\right)^2
    \left[
    \frac{H}{r^2}
    -
    \frac{1}{2}\frac{d}{dr}\left(\frac{H}{r}\right)
    \right]
    \Biggr]
    \nonumber \\
+
&&
    \frac{A_2L_2}{2\pi r} 
     \Biggr[
     \frac{d}{dr}\left(\frac{H}{r}\right)
     -
    \beta_2\left(\frac{r_{\rm in}}{r}\right)^2
    \left[
    \frac{H}{r^2}
    -
    \frac{1}{2}\frac{d}{dr}\left(\frac{H}{r}\right)
    \right]
    \Biggr]
    =
    2\sigma{T^4},
    \label{eq:eneeq}
\end{eqnarray}
where we used equations~(\ref{eq:qrad}), (\ref{eq:qvis}), and (\ref{eq:qirrtot}) for the derivation. In equation~(\ref{eq:eneeq}), $H=H(r)$ is the scale height of the CBD, $L_1$ and $L_2$ are the bolometric luminosities of the primary and secondary black holes, respectively, given by the equation~(\ref{eq:lumii}) as
\begin{equation}
    L_1=\frac{1}{6}\dot{M}_1c^2,\,\,
    L_2=\frac{1}{6}\dot{M}_2c^2,
\end{equation}
$\beta_1$ and $\beta_2$ are given in equation~(\ref{eq:beta-param}), and
\begin{eqnarray}
    r_{\rm in}
    &&
    =C_{\rm gap}a
    \nonumber \\
    &&
     =6.0\times10^{10}\,{\rm cm}
     \left(\frac{C_{\rm gap}}{2}\right)
     \left(\frac{a_0}{1000}\right)
     \left(\frac{M}{100\,M_\odot}\right)
\end{eqnarray}
is the CBD inner-edge radius with $1.6\lesssim C_{\rm gap}\lesssim4$ \citep{artymowicz_dynamics_1994,2023ApJ...949L..30D}, where $C_{\rm gap}=2$ is adopted as a fiducial value throughout this paper.

The CBD is also assumed to be in hydrostatic equilibrium for the direction perpendicular to the disk plane:
\begin{eqnarray}
    c_{\rm s}=\Omega{H},
    \label{eq:hydrostaticeq}
\end{eqnarray}
where $\Omega=\sqrt{GM/r^3}$ is the Keplerian angular frequency and
\begin{eqnarray}
c_{\rm s}=\sqrt{\frac{R_{\rm g}}{\mu}T_{\rm c}}
\label{eq:soundspeed}
\end{eqnarray}
is the sound speed with the molecular weight for an ionized plasma with solar abundances being $\mu=0.615$ and the gas constant $R_{\rm g}$, giving the CBD mid-plane temperature:
\begin{eqnarray}
T_{\rm c}=\frac{\mu}{R_{\rm g}}
\left(\frac{H}{r}\right)^2
\frac{GM}{r}.
\label{eq:mid-plane-temp}
\end{eqnarray}
In what follows, we assume that the CBD mid-plane temperature is approximately equal to the surface temperature, i.e., $T_{\rm c}\approx T$. The validity of this assumption will be discussed later in the Discussion section.

%
\section{Method} 
\label{sec:method}
%

Solving the energy equation yields the disk aspect ratio $H/r$, which can then be substituted into the hydrostatic equation to finally obtain the radial distribution of the CBD temperature. In the following, we describe the method, divided into two ways: one is to approximate the energy equation to obtain an analytical solution, and the other is to solve the energy equation numerically. To solve the energy equation prospectively, we introduce the following dimensionless variables:
\begin{eqnarray}
    \xi&\equiv&\frac{r}{r_{\rm in}},
    \nonumber \\
    Y&\equiv&\frac{H}{r}.
    \label{eq:dimlessparams}
\end{eqnarray}

%
\subsection{Analytical solutions}
%

The energy equation is approximated into two separate equations, depending on the dominant heating mechanism: $Q_{\rm vis}=Q_{\rm rad}$ for $Q_{\rm vis}\gg Q_{\rm irr}$ and $Q_{\rm irr}=Q_{\rm rad}$ for $Q_{\rm vis}\ll Q_{\rm irr}$. The former case gives the temperature profile in the viscous heating dominated region directly without using equation~(\ref{eq:hydrostaticeq}):
\begin{eqnarray}
T=T_{\rm in,vis}\,\xi^{-3/4},
\label{eq:tvis}
\end{eqnarray}
where
\begin{eqnarray}
    T_{\rm in,vis}
    &&
    =
     \left( \frac{3}{8\pi}\frac{GM\dot{M}}{\sigma r_{\rm in}^3} \right)^{1/4}
     \nonumber \\
     &&
     \sim
     3.7\times10^{4}\,{\rm K}
     \left(\frac{M}{100\,M_\odot}\right)^{-1/4}
     \left(\frac{\dot{m}}{1.0}\right)^{1/4}
     \left(\frac{C_{\rm gap}}{2}\right)^{-3/4}
     \left(\frac{a_0}{1000}\right)^{-3/4}.
     \label{eq:tvisin}
\end{eqnarray}
Next, we consider the latter, irradiation-heating dominated case. Applying equation~(\ref{eq:dimlessparams}) to equations~(\ref{eq:qirr1}) and (\ref{eq:qirr2}) gives irradiation heating rates with dimensionless variables:
\begin{eqnarray}
\langle{Q_{\rm irr,1}}\rangle
    &&
    =\frac{A_1L_1}{2\pi r_{\rm in}^2}\frac{1}{\xi} \left[\frac{dY}{d\xi}-\frac{\beta_1}{\xi^2}\left(\frac{Y}{\xi}-\frac{1}{2}\frac{dY}{d\xi}\right)\right],
\nonumber \\
\langle{Q_{\rm irr,2}}\rangle
    &&
=\frac{A_2L_2}{2\pi r_{\rm in}^2}\frac{1}{\xi} \left[\frac{dY}{d\xi}-\frac{\beta_2}{\xi^2}\left(\frac{Y}{\xi}-\frac{1}{2}\frac{dY}{d\xi}\right)\right],
\nonumber
\end{eqnarray}
where $\beta_1$ and $\beta_2$ are given by equation~(\ref{eq:beta-param}). Then, total irradiation heating rate is given as the sum of $ \langle{Q_{\rm irr,1}}\rangle$ and $ \langle{Q_{\rm irr,2}}\rangle$ by 
\begin{equation}
Q_{\rm irr}
    =\frac{A_1L_1}{2\pi r_{\rm in}^2}
    \frac{1}{\xi}\Biggr[ (1+Q_{12})\frac{dY}{d\xi}-
    \frac{\beta_1+Q_{12}\beta_2}{\xi^2}
    \left(\frac{Y}{\xi} -\frac{1}{2}\frac{dY}{d\xi}
    \right)\Biggr],
    \nonumber \\
\label{eq:dimlessqirrtot}
\end{equation}
where $Q_{12}=A_2L_2/(A_1L_1)$. Since $A_1$ and $A_2$ are unlikely to have much different values for two minidisks of the similar physical state, we assume that $A_1=A_2=A$, resulting in $Q_{12}=q$.

Equation~(\ref{eq:mid-plane-temp}) is rewritten with equation~(\ref{eq:dimlessparams}) as
\begin{equation}
    T=T_0\frac{Y^2}{\xi},
    \label{eq:hst}
\end{equation}
where we define the dynamical temperature at $r_{\rm in}$ as 
\begin{eqnarray}
    T_0
    &&
    =
    \frac{\mu}{R_g}\frac{GM}{r_{\rm in}}    \nonumber \\
    &&
    \sim 
    1.66\times10^9\,{\rm K}
    \left(\frac{C_{\rm gap}}{2}\right)^{-1}
    \left(\frac{a_0}{1000}\right)^{-1}.
    \nonumber 
    \label{eq:t0}
\end{eqnarray}
Combining equation~(\ref{eq:qrad}) with equation~(\ref{eq:hst}), we write the radiative cooling rate as
\begin{eqnarray}
Q_{\rm rad} =2\sigma{T_0^4}\frac{Y^8}{\xi^4}.
\label{eq:dimlessqrad}
\end{eqnarray}
Equating equation~(\ref{eq:dimlessqirrtot}) with equation~(\ref{eq:dimlessqrad}) gives the following differential equation: 
\begin{eqnarray}
    \frac{dY}{d\xi}
    =\left[\alpha Y^8+\beta{Y} \right]
    \left[\xi^3+\frac{\beta}{2}\xi\right]^{-1},
\label{eq:dydx1}
\end{eqnarray}
where we introduce the following two parameters:
\begin{eqnarray}
    \alpha
    &&
    =
    \frac{1}{1+Q_{12}}
    \frac{1}{A}
    \frac{L_{0}}{L_1}
    =
    \frac{1}{A}\frac{L_0}{L_1+L_2}
    ,
    \label{eq:alpha}
    \\
    \beta
    &&
    =
    \frac{\beta_1+\beta_2Q_{12}}{1+Q_{12}}
     =
    \beta_1\frac{L_1}{L_1+L_2}
    +
    \beta_2\frac{L_2}{L_1+L_2}
    \label{eq:beta}
\end{eqnarray}
with the normalization luminosity defined by
\begin{eqnarray}
L_{0}
&&
\equiv 4\pi r_{\rm in}^2
    \sigma{T}_{0}^4
    \nonumber \\
    &&
    \sim 1.9\times10^{55}\,{\rm ergs^{-1}}
    \left(\frac{M}{100\,M_\odot}\right)^{2}
    \left(\frac{C_{\rm gap}}{2}\right)^{-2}
    \left(\frac{a_0}{1000}\right)^{-2}.
    \nonumber \\
    \label{eq:l0} 
\end{eqnarray}
Here, $\alpha$ is the ratio of the blackbody luminosity at the inner edge of the CBD to the total irradiation luminosity from the two minidisks, whereas $\beta$ is an average of $\beta_1$ and $\beta_2$, weighted by the ratio of the total irradiation to respective irradiation luminosities.

Equation~(\ref{eq:dydx1}) can be easily integrated analytically (see Appendix~\ref{app:anasol} for details), and its special solution is given by equation~(\ref{eq:yanalsol2}). Now, it is noted that $\beta/\xi^2\ll1$, since the region where irradiation heating dominates is the outer part of the disk. 
Expanding equation~(\ref{eq:yanalsol2}) in a Taylor series of $\beta/\xi^2$ and keeping up to the first-order terms yields the following approximate solution:
\begin{eqnarray}
    Y
    &\approx&
    \left(\frac{2}{7\alpha}\right)^{1/7}
    \xi^{2/7}\left(1-\frac{3}{14}\frac{\beta}{\xi^2}
    \right).
    \label{eq:appsol1}
\end{eqnarray}
This solution reduces to a simple power-law solution of $\xi$ in the $\beta=0$ case: 
\begin{eqnarray}
    Y=\left(\frac{2}{7\alpha}\right)^{1/7}\xi^{2/7},
    \label{eq:appsol2}
\end{eqnarray}
which is consistent with the solution of the single black hole case  \citep{1981PASJ...33..365H,1993PASJ...45..443S}, except for the value of $\alpha$.

Substituting equation~(\ref{eq:appsol1}) into equation~(\ref{eq:hst}) gives
\begin{eqnarray}
    T= T_0\left(\frac{2}{7\alpha}\right)^{2/7}
    \xi^{-3/7}
    \left(1-\frac{3}{14}\frac{\beta}{\xi^2}
    \right)^2.
    \label{eq:tirr}
\end{eqnarray}
This is the radial dependence of the CBD temperature in the irradiation-dominated region.
%
\subsection{Numerical solutions}
%
Next, we derive the radial profile of the CBD temperature numerically. With dimensionless variables, the energy balance equation $Q_{\rm vis} + Q_{\rm irr}=Q_{\rm rad}$ is 
expressed as  
\begin{eqnarray}
    &&
    \frac{3}{4\pi}\frac{GM\dot{M}}{r_{\rm in}^3}\frac{1}{\xi^3}
    +\frac{AL_1}{2\pi r_{\rm in}^2}\frac{1}{\xi}
    \left[ (1+Q_{12})\frac{dY}{d\xi}-
    \frac{\beta_1+Q_{12}\beta_2}{\xi^2}
    \left(\frac{Y}{\xi} -\frac{1}{2}\frac{dY}{d\xi}\right)\right]
    =2\sigma{T_{0}^4}\frac{Y^8}{\xi^4}.
    \label{eq:fullene}
    \end{eqnarray}
This leads to the differential equation to determine the evolution of the disk aspect ratio as    
\begin{eqnarray}
    \frac{dY}{d\xi}
    =\left[\alpha\frac{Y^8}{\xi^3}+\beta\frac{Y}{\xi^3} -\gamma\frac{1}{\xi^2}\right]
    \left[1+\frac{\beta}{2}\frac{1}{\xi^2}\right]^{-1},
    \label{eq:fulldiff}
\end{eqnarray}
where $\gamma$ is defined as
\begin{eqnarray}
\gamma
&&
=\frac{3}{2}\frac{1}{AL_1}
    \frac{1}{1+Q_{12}}
    \frac{GM\dot{M}}{r_{\rm in}}
    =\frac{9}{2}\frac{1}{A}
    \frac{1}{C_{\rm gap}a_0}
    \nonumber \\
    &&
\sim
2.3\times10^{-2}
\left(\frac{A}{0.1}\right)^{-1}
\left(\frac{C_{\rm gap}}{2}\right)^{-1}
\left(\frac{a_0}{1000}\right)^{-1}.
\label{eq:gamma}
\end{eqnarray}
We solve equation~(\ref{eq:fulldiff}) numerically with the outer boundary condition:
\begin{eqnarray}
Y_{\rm out}=
\left(\frac{2}{7\alpha}\right)^{1/7}
\xi_{\rm out}^{2/7}\left(1-\frac{3}{14}\frac{\beta}{\xi_{\rm out}^2}\right),
\end{eqnarray}
where equation~(\ref{eq:appsol1}) is adopted at $\xi_{\rm out}=r_{\rm out}/r_{\rm in}$ with $r_{\rm out}$ being the outer radius of the CBD. The radial temperature profile is then obtained by substituting the numerical solution of $Y$ into equation~(\ref{eq:hst}).

To sum up at the end, our model has seven parameters: $M$, $\dot{m}$, $a_0$, $C_{\rm gap}$, $q$, $A$, and $\xi_{\rm out}$. Among them, the following four parameters are fixed through the paper as $\dot{m}=1$, $a_0=1000$, $C_{\rm gap}=2$, and $A=0.1$. The dependencies of the CBD temperature on the remaining three parameters and spectra are examined in the next section.

%
\section{Results}
%
In this section, we compare the analytical and numerical solutions for the radial temperature distribution of the CBD and discuss how it depends on some parameters.
%
\subsection{
Comparison of circumbinary disk temperature between analytical and numerical solutions
}
%

\begin{figure*}[http]
    \centering
\includegraphics[scale=0.58]{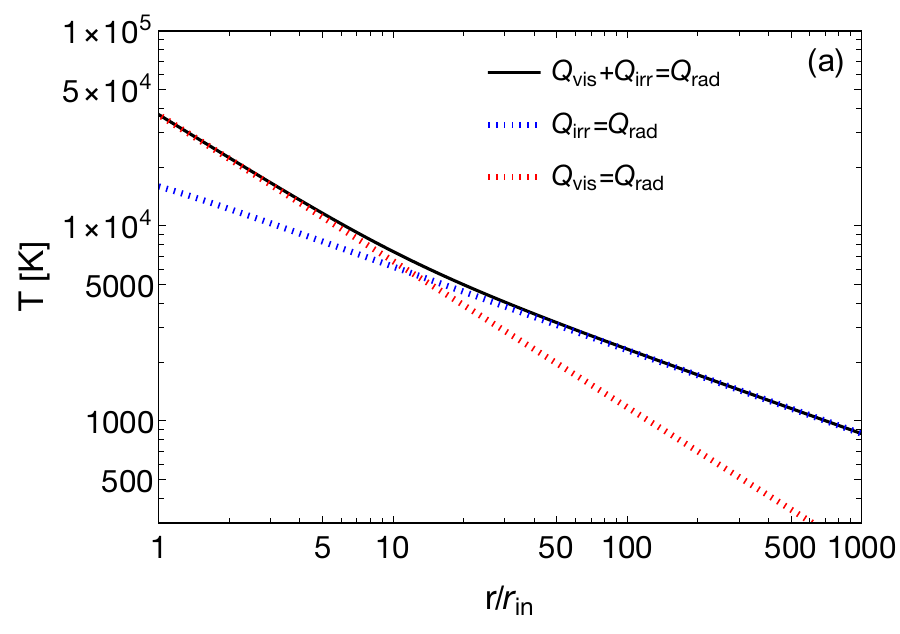}
\includegraphics[scale=0.58]{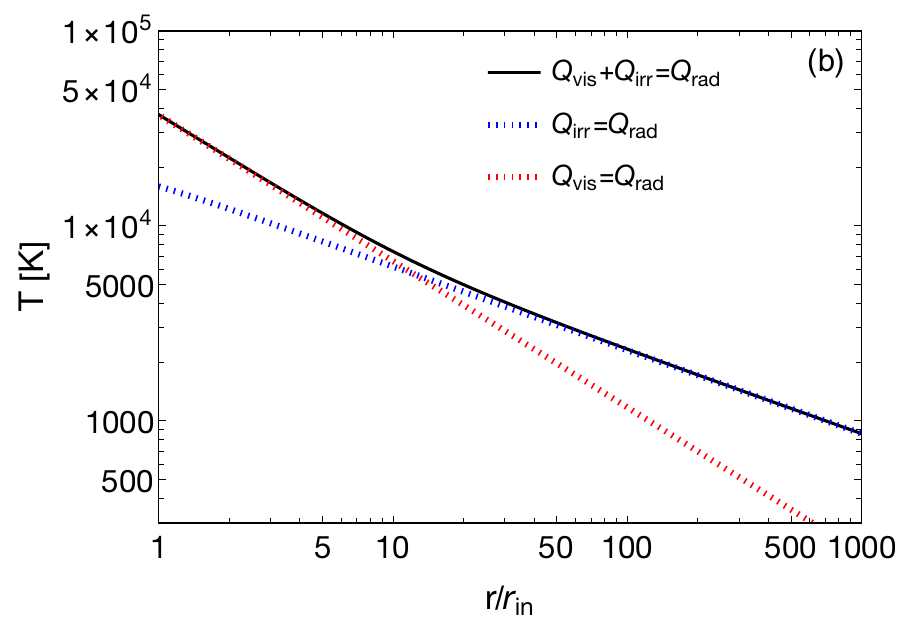}
\includegraphics[scale=0.58]{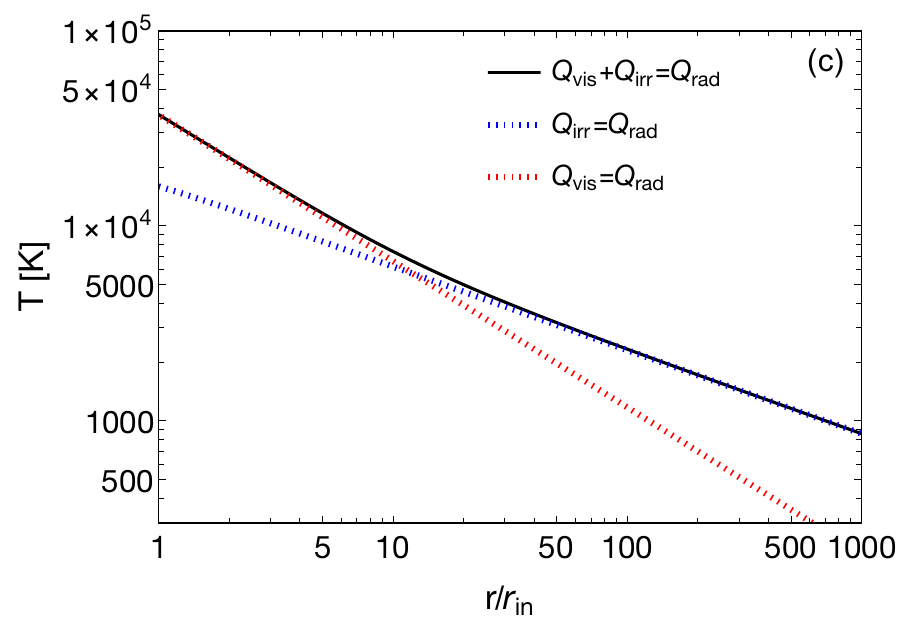}
\includegraphics[scale=0.58]{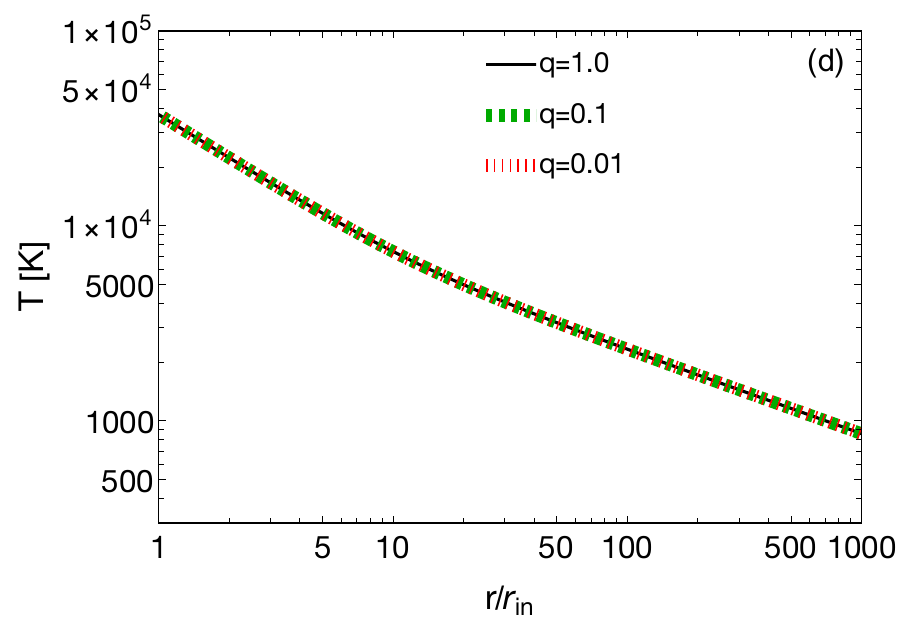}
    \caption{
Comparison of the radial temperature profiles between the analytical and numerical solutions for the different mass ratios. The black solid line denotes the numerical solution for the energy equation~(\ref{eq:fullene}), while the red and blue dashed lines are the analytical solutions for $Q_{\rm vis}=Q_{\rm rad}$ and $Q_{\rm irr}=Q_{\rm rad}$, which are given by equations~(\ref{eq:tvis}) and (\ref{eq:tirr}). Panels~(a), (b), and (c) represent the $q=1.0$, $q=0.1$, and $q=0.01$ cases, respectively, while panel (d) compares the numerical solutions of these cases. 
    }
   \label{fig:temp}
\end{figure*}

Figure~\ref{fig:temp} compares the radial temperature profiles between the analytical and numerical solutions for three different mass ratios. Panels~(a), (b), and (c) represent the radial profiles of the CBD temperature for the cases of $q=1.0$, $q=0.1$, and $q=0.01$, respectively, while panel (d) shows the comparison between the numerical solutions of these cases.

Panels (a)-(c) of Figure~\ref{fig:temp} exhibit that the analytical and numerical solutions are in good agreement. It can also be seen from panel (d) that there is almost no dependence on the mass ratio. This is because the CBD temperature due to viscous heating depends only on the total BH mass and not on the mass ratio. In addition, $\alpha$ in equation~(\ref{eq:alpha}) is also independent of the mass ratio. 
In contrast, the binarity effect of irradiation heating appears in $\beta$ of equation~(\ref{eq:beta}). 
However, panel (d) indicates that the CBD temperature negligibly depends on the mass ratio because of $\beta\ll1$.

%
\subsection{Cicumbinary disk spectra}
%
Since the CBD is highly optically thick in the vertical direction, its surface locally radiates the blackbody radiation with the spectral intensity:
\begin{eqnarray}
    I_\nu=\frac{2h}{c^2}\frac{\nu^3}{\exp(h\nu/kT)-1},
    \nonumber 
\end{eqnarray}
where $h$ is the Planck constant, $k$ is the Boltzmann constant, and $\nu$ is a frequency. The flux density to be emitted from the whole of the CBD is then given by \citep{kato_black-hole_2008}
\begin{eqnarray}
S_{\nu}
=
\int I_\nu\,
d\Omega
=
4\pi
\frac{h}{c^2}
\frac{\cos{\delta}}{D^2}
\nu^3
\int_{r_{\rm in}}^{r_{\rm out}}
\,
\frac{r}{e^{h\nu/(kT)}-1}
\,dr,
\nonumber \\
\label{eq:snu}
\end{eqnarray}
where $\delta$ is an inclination angle of the disk and $D$ is the distance between the source and the earth. In the following we adopt $\delta=0$ unless otherwise noted.

In our analytical models, since the disk's inner part is dominated by viscous heating, the disk temperature obeys $T/T_{\rm in}=(r/r_{\rm in})^{-3/4}$ in viscous-heating dominated region. In contrast, the disk's outer part is dominated by irradiation heating. Since $Q_{\rm vis}=Q_{\rm irr}$ gives the boundary radius between the two regions, we estimate it as
\begin{eqnarray}
    \xi_{\rm b}
    &&
    =
    \left(\frac{7\alpha}{2}\right)^{1/9}
    \left(
    \frac{21}{4}
    \frac{\alpha}{L_0}
    \frac{GM\dot{M}}{r_{\rm in}}
    \right)^{7/9}
    \nonumber \\
    &&
    \sim
    1.2\times10 \left(\frac{\dot{m}}{1.0}\right)^{-1/9}
    \left(\frac{M}{100\,M_\odot}\right)^{1/9}
    \left(\frac{A}{0.1}\right)^{-8/9}
    \left(\frac{C_{\rm gap}}{2}\right)^{-1}
    \left(\frac{a_0}{1000}\right)^{-1}
    ,
\end{eqnarray}
equations~(\ref{eq:hydrostaticeq}), (\ref{eq:dimlessparams}), (\ref{eq:tvis}), and (\ref{eq:appsol2}) are used for the derivation.
The boundary radius $\xi_{\rm b}$ is independent of the mass ratio and is insensitive to both the binary mass and the mass accretion rate. We also note that $\xi_{\rm b}$ is a dozen times larger than the inner edge radius of the CBD. Combining equations~(\ref{eq:tvis}), (\ref{eq:hst}), and (\ref{eq:appsol2}) gives $T/T_{\rm b}=(r/r_{\rm b})^{-3/7}$ as the CBD temperature of the irradiation-heating dominated region, where 
\begin{eqnarray}
    T_{\rm b}
    &&
    =
    \left(\frac{7\alpha}{2}\right)^{-1/12}\left(\frac{3}{8\pi}\frac{GM\dot{M}}{\sigma r_{\rm in}^3}\right)^{1/4}
        \left(
    \frac{21}{4}
    \frac{\alpha}{L_0}
    \frac{GM\dot{M}}{r_{\rm in}}
    \right)^{-7/12}
    \nonumber \\
    &&
    \sim 5.7\times10^3\,{\rm K}
    \left(\frac{\dot{m}}{1.0}\right)^{1/3}
    \left(\frac{M}{100\,M_\odot}\right)^{-1/3}
    \left(\frac{A}{0.1}\right)^{2/3}.
    \nonumber \\
\end{eqnarray} 
We note that the temperature at $\xi_{\rm b}$ is independent of the inner edge radius of the CBD.
 
The disk spectrum of the viscous-heating dominated region is given by
\begin{equation}
S_{\nu,\rm vis}
=
\frac{16\pi}{3}
\frac{h}{c^2}
\left(
\frac{r_{\rm in}}{D}
\right)^2
\left(
\frac{kT_{\rm in}}{h\nu}
\right)^{8/3}
\nu^3
\int_{\zeta_{\rm in}}^{\zeta_{\rm b}}
\,
\frac{\zeta^{5/3}}{e^{\zeta}-1}
d\zeta,
\label{eq:snuvis}
\end{equation}
where $\zeta=h\nu/kT$, $\zeta_{\rm in}=h\nu/kT_{\rm in}$, and $\zeta_{\rm b}=h\nu/kT_{\rm b}$.
On the other hand, the disk spectrum of the irradiation-heating dominated region is given by
\begin{eqnarray}
S_{\nu,\rm irr}
=
\frac{28\pi}{3}
\frac{h}{c^2}
\left(
\frac{r_{\rm in}}{D}
\right)^2
\left(
\frac{kT_{\rm in}}{h\nu}
\right)^{14/3}
\nu^3
\int_{\zeta_{\rm b}}^{\zeta_{\rm out}}
\,
\frac{\zeta^{11/3}}{e^{\zeta}-1}
d\zeta,
\nonumber 
\\
\label{eq:snuirr}
\end{eqnarray}
where $\zeta_{\rm out}=h\nu/kT_{\rm out}$ and $T_{\rm out}=T_{\rm b}\,(\xi_{\rm out}/\xi_{\rm b})^{-3/7}$.

\begin{figure*}[http]
    \centering
\includegraphics[scale=0.58]{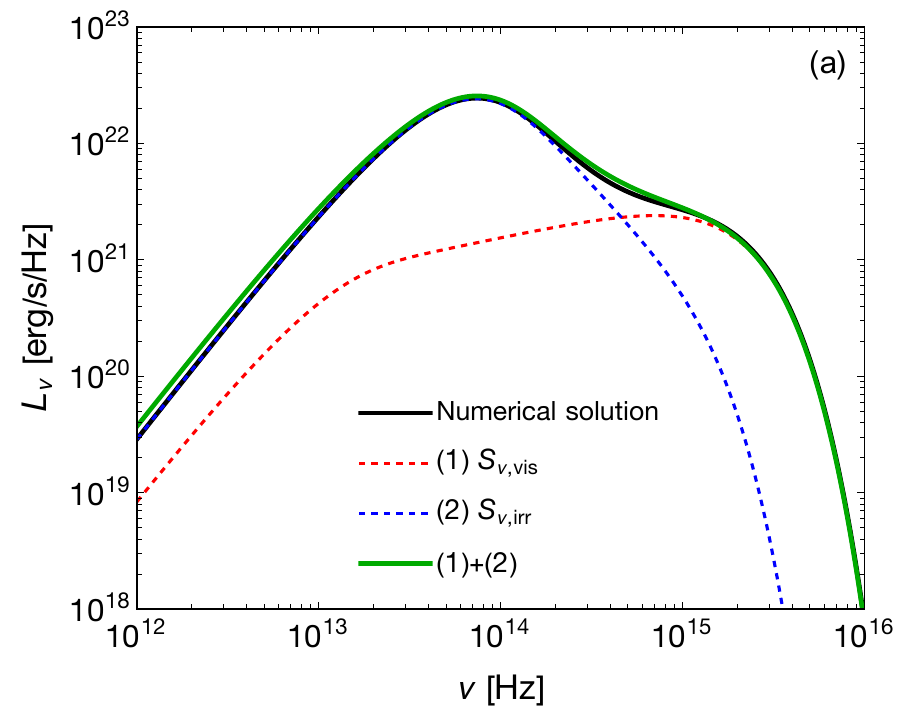}
\includegraphics[scale=0.58]{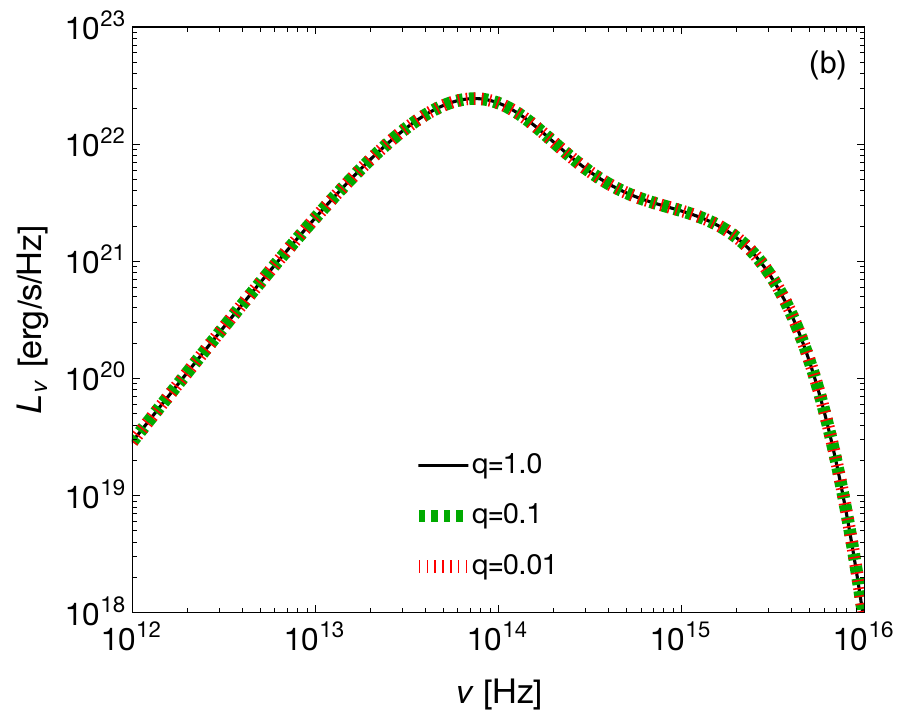}
\includegraphics[scale=0.58]{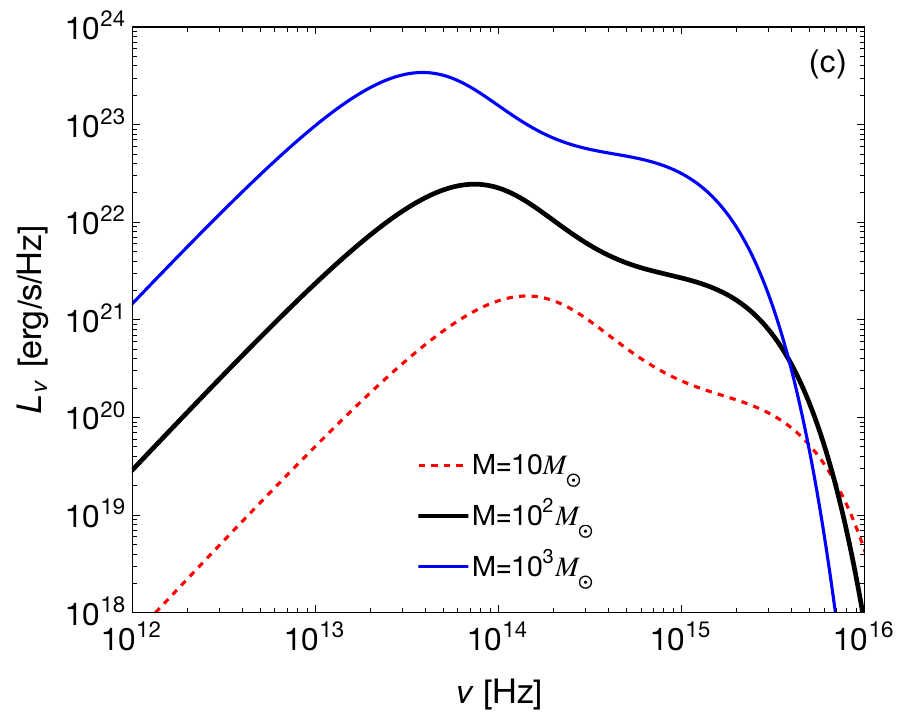}
\includegraphics[scale=0.58]{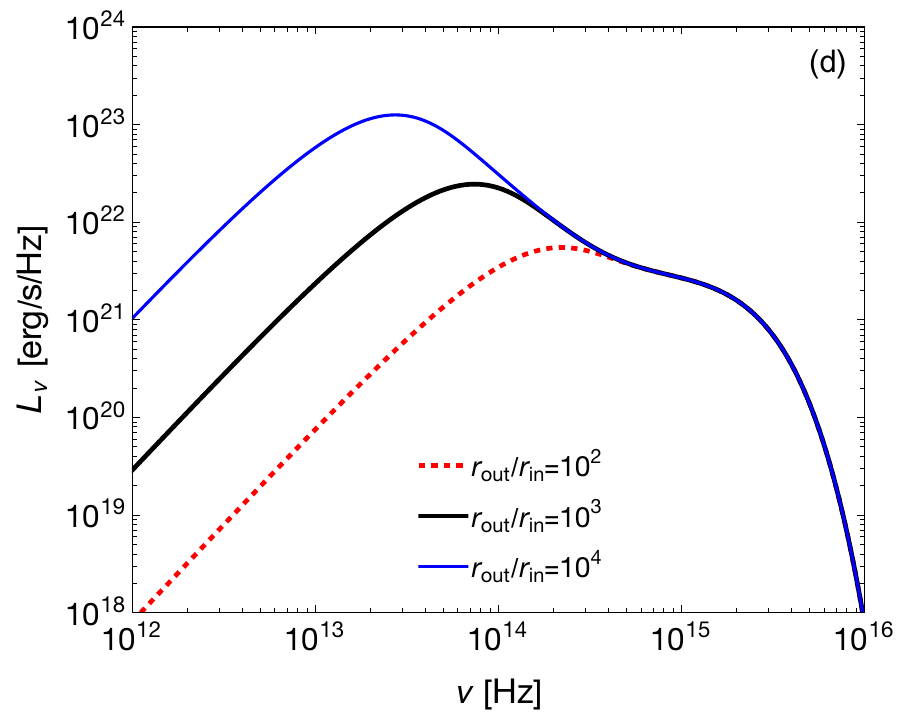}
    \caption{
Spectral energy distributions of the CBD. The vertical and horizontal axes represent spectral luminosity and frequency, respectively. Both axes are measured on a logarithmic scale. For our fiducial model, the binary mass is $10^2\,M_\odot$, the mass ratio is $q=1.0$, and the outer boundary radius is $r_{\rm out}=10^3\,r_{\rm in}$. In all four panels, the solid thick black line represents the fiducial model. 
Panel (a) shows the comparison between the analytical and numerical solutions. The red and blue dashed lines denote the analytical solutions: (1) $Q_{\rm vis}=Q_{\rm rad}$ and (2) $Q_{\rm irr}=Q_{\rm rad}$ cases, respectively. While the black solid line represents the numerical solutions, the green solid thick line represents the sum of (1) and (2). Panel (b) display model spectra for three different mass ratios. The green dashed and red dotted lines denote the $q=0.1$ and $q=0.01$ cases, respectively. Panel (c) demonstrates the effect of the binary mass. The blue solid and red dashed lines represent the $M=10^3\,M_\odot$ and $M=10\,M_\odot$ cases, respectively. Panel (d) shows the spectral dependence on the outer radius of the CBD. The blue solid and red dashed lines represent the $r_{\rm out}=10^4\,r_{\rm in}$ and $r_{\rm out}=100\,r_{\rm in}$ cases, respectively.
    }
   \label{fig:CBD}
\end{figure*}

Figure~\ref{fig:CBD} shows the spectral energy distributions of the CBD. For our fiducial model, we adopt that the binary mass is $10^2\,M_\odot$, the mass ratio is $q=1.0$, and the outer boundary radius is $r_{\rm out}=10^3\,r_{\rm in}$. In all four panels, the thick solid black line represents the fiducial model. Panel (a) compares the analytical solutions $4\pi{D^2}S_{\nu,\rm vis}$, $4\pi{D^2}S_{\nu,\rm irr}$, and $4\pi{D^2}(S_{\nu,\rm vis}+S_{\nu,\rm irr})$ with the numerical solutions $4\pi{D^2}S_\nu$. It is noted from the panel that the CBD spectrum generally has two peaks, one peak arizing from the viscous-heating dominated region and the other from the irradiation-heating dominated region. We also find that the numerical solution is in good agreement with the analytical solution.

Panel (b) displays the model spectra for three different mass ratios, demonstrating that the CBD spectra hardly depend on the mass ratio as predicted in panel (b) of Figure~\ref{fig:temp}. Panel (c) demonstrates the dependence of the spectral luminosity on the binary mass. We note that the spectral luminosity increases with increase in the black hole mass. The spectral luminosity also depends on the outer radius of the CBD. Panel (d) shows the effect of the outer radius. The lower-energy peak becomes higher as the outer radius increases, because of the increase in the emitting area.

%
%

%
\subsection{Triple disk spectra}
%
This section describes the spectral energy distributions of the triple disk system consisting of the two minidisks and the CBD surrounding them. The spectral luminosities of the CPD and the CSD are calculated by equation (\ref{eq:snu}) as
\begin{eqnarray}
L_{\nu,i}=
16\pi^2
\frac{h}{c^2}
\nu^3
\int_{r_{\rm in,i}}^{r_{\rm out,i}}
\,
\frac{r}{e^{h\nu/(kT_i)}-1}
\,dr,
\label{eq:lnui}
\end{eqnarray}
where the inner boundary radius is set to the ISCO radius of each black hole, i.e., $r_{{\rm in},i}=6GM_i/c^2$ with $i=1$ for the CPD and $i=2$ for the CSD. We also assume that the disk outer disk radius is equal to the Roche radius of each black hole \citep{1983ApJ...268..368E}:
\begin{eqnarray}
    r_{{\rm out},1},
    &&
    =\frac{0.49q^{-2/3}a}{0.6q^{-2/3}+\ln{(1+q^{-1/3})}},
    \nonumber \\
    r_{\rm out,2}
    &&
    =\frac{0.49q^{2/3}a}{0.6q^{2/3}+\ln{(1+q^{1/3})}},
    \nonumber
    \end{eqnarray}
and the temperature of each disk is calculated by 
    \begin{eqnarray}
    T_i
    &&
    =\left(
    \frac{8}{3\pi}
    \frac{GM_i\dot{M}_i}{r_{{\rm in},i}}\right)^{1/4} 
    \left(\frac{r}{r_{{\rm in},i}}\right)^{-3/4}.
\end{eqnarray}

\begin{figure*}[http]
    \centering
\includegraphics[scale=0.58]{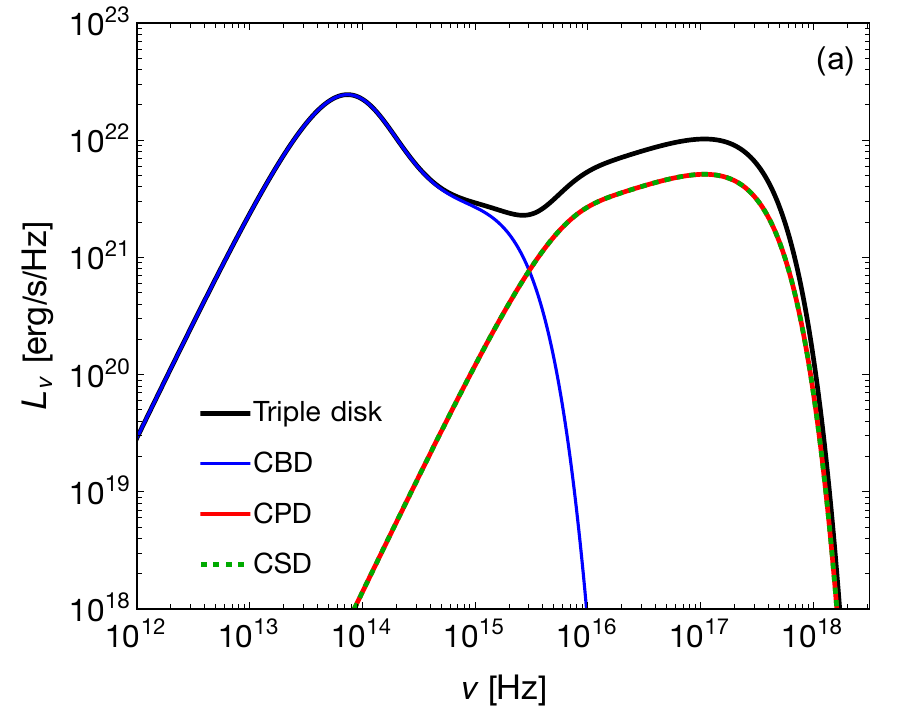}
\includegraphics[scale=0.58]{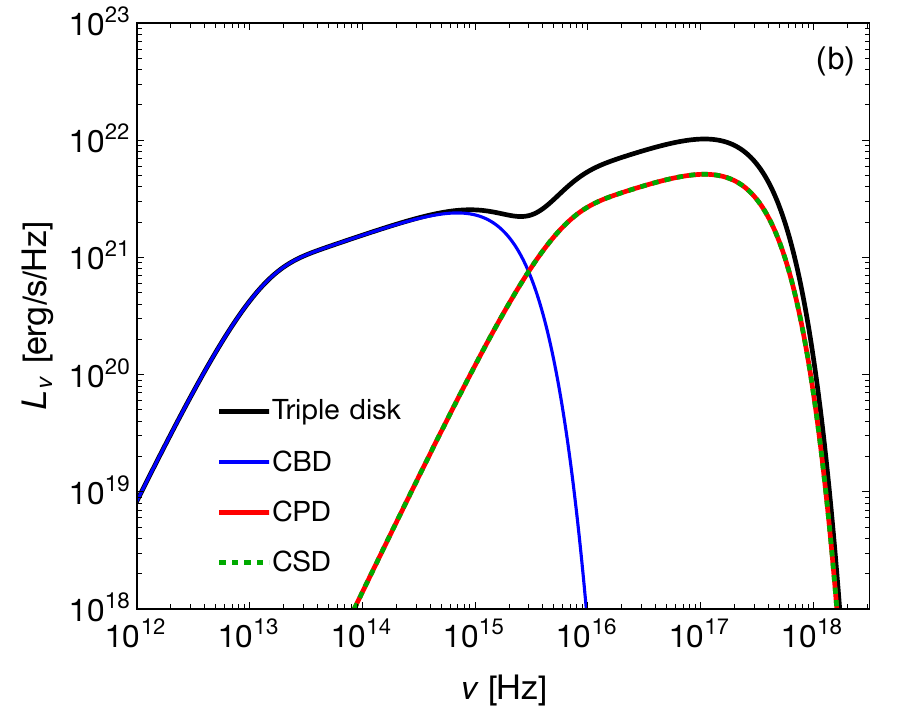}
    \caption{
 Comparison of the spectral energy distributions of the triple disk, consisting of the circum-primary disk (CPD), the circum-secondary disk (CSD) and the CBD surrounding them, between two different cases. One case is the triple disk spectra with the CBD with irradiation as shown in panel (a), while the other case is the triple disk spectra with the  CBD without irradiation as shown in panel (b). For both CBDs, the same parameters as in Figure~\ref{fig:CBD} are adopted. The vertical and horizontal axes represent the spectral luminosity and frequency, respectively. Both axes are measured on a logarithmic scale.  In panels (a) and (b), the blue solid lines denote the CBD spectrum with and without irradiation, respectively. In both panels, the red and green dashed lines denote the CPD and CSD spectra, respectively, while the black solid line shows the combined spectra of each triple disk.
    }
   \label{fig:triple}
\end{figure*}

Figure~\ref{fig:triple} compares the spectral energy distributions of the triple disk between two different cases. Panel (a) represents the triple disk spectra with the CBD with irradiation, while panel (b) shows the triple disk spectra with the CBD without irradiation, where the blue solid lines denote the CBD spectrum. In deriving these spectra, the same parameters are adopted as in Figure~\ref{fig:CBD}. It is also noted from panel (a) that the spectral luminosity increases with frequency as $\nu^{1/3}$ toward the high-frequency peak, while it decreases with frequency as $\nu^{-5/3}$ in the NIR to optical wavebands.

\begin{figure*}[http]
    \centering
\includegraphics[scale=0.58]{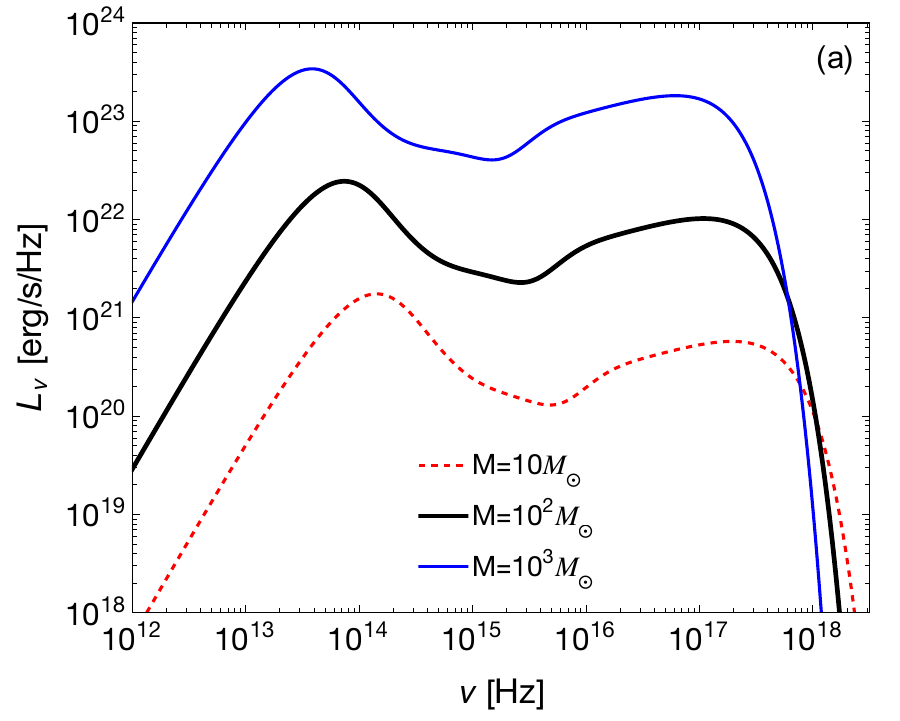}
\includegraphics[scale=0.58]{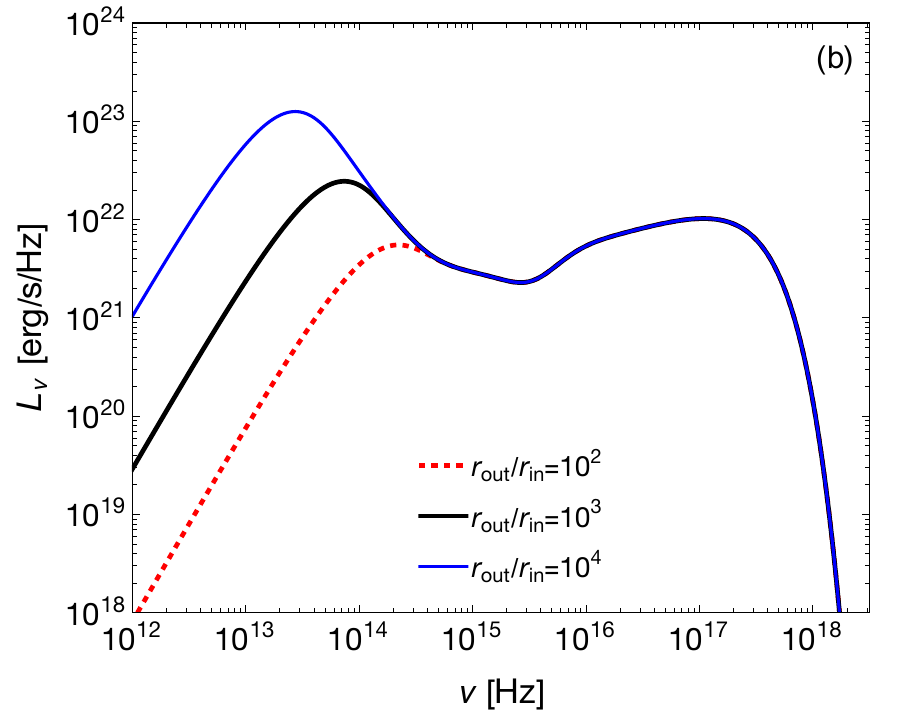}
    \caption{
    Spectral energy distributions of the triple disk with different parameters. Both axes are measured on a logarithmic scale. The fiducial model has the same parameters as in the panel (a) of Figure~\ref{fig:triple}. In all panels, the solid thick black line represents the fiducial model. In panel (a), the blue solid and red dashed lines represent the $M=10^3\,M_\odot$ and $M=10\,M_\odot$ cases, respectively. In panel (b), the blue solid and red dashed lines represent the $r_{\rm out}=10^4\,r_{\rm in}$ and $r_{\rm out}=100\,r_{\rm in}$ cases, respectively.
    }
   \label{fig:triple2}
\end{figure*}

Figure~\ref{fig:triple2} shows the triple disk spectra for the different black hole masses and outer boundary radii. Panel~(a) of Figure~\ref{fig:triple2} shows how the triple disk spectra depend on the black hole mass, indicating that the luminosity is higher as the black hole mass increases. Panel~(b) shows how the spectra depend on the outer boundary radius. From panel~(b) it can be seen that only the low frequency peak of the spectral luminosity increases as $r_{\rm out}$ becomes larger. This is simply because the area over which the low-frequency photons are emitted is larger.

%
%

%
%
%
\subsection{Observational implications}
%
In this section we discuss the observability of our model spectra by comparing them with the flux limits of the Hyper Suprime-Cam (HSC) Subaru Strategic Program survey and the James Webb Space Telescope (JWST) in the IR to optical wavelength range, and {\it Swift}/XRT in the X-ray waveband. 
The sensitivity data of these instruments are taken from their web pages.

\begin{figure*}[ht!]
\centering
\includegraphics[scale = 0.65]{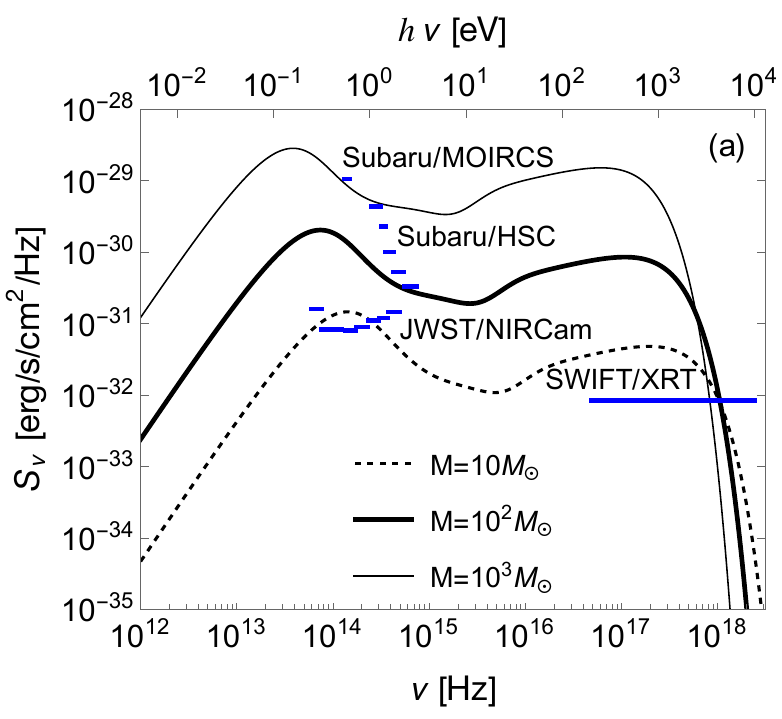}
\includegraphics[scale = 0.65]{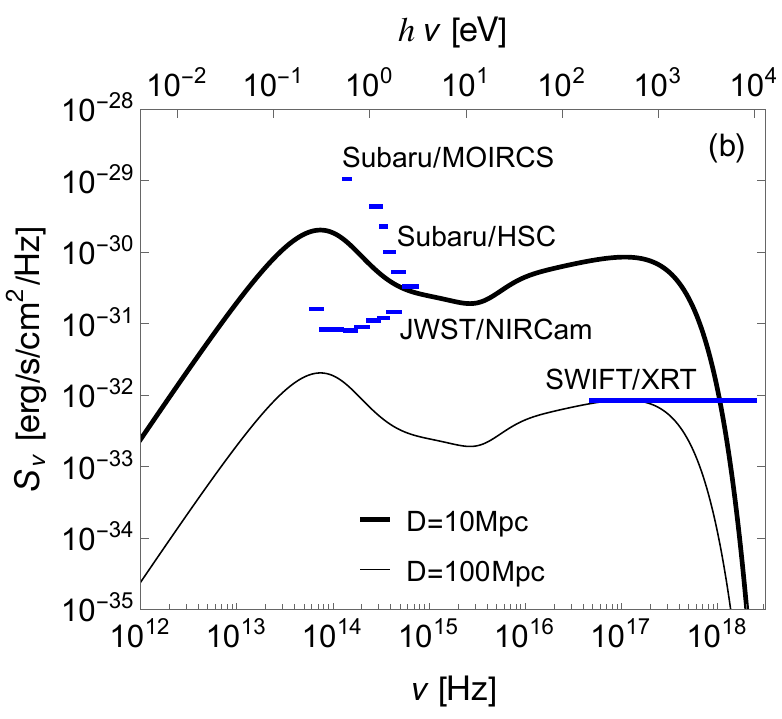}
\includegraphics[scale = 0.65]{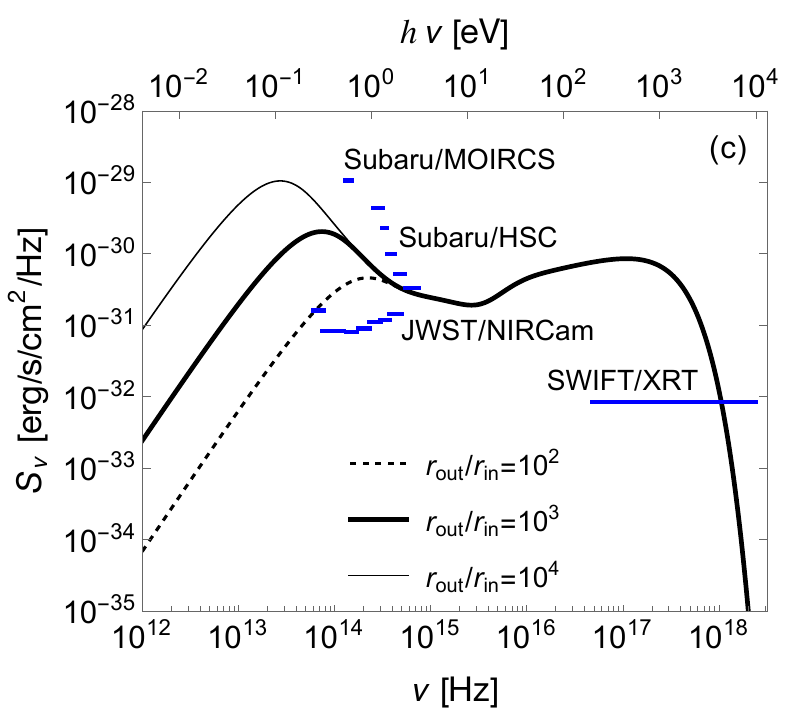}
\caption{
Comparison of Subaru, JWST, and {\it Swift}/XRT flux limits and theoretical triple disk spectra with different BH masses (panel a), distances (panel b), and outer boundary radii (panel c). The fiducial values for the BH mass, distance, and outer boundary radius are $M_{\rm bh}=100M_\odot$, $D=10\,{\rm Mpc}$, and $r_{\rm out}=1000r_{\rm in}$, respectively.
Shown on top of the spectra are the sensitivities of the JWST/NIRCam filters (F444W, F332W2, F200W, F150W, F115W, F090W, and F070W from left to right), Subaru/MOIRCS, and Subaru/HSC filters (Y-, z-, i2-, r2-, and g-bands from left to right) in optical-IR, and {\it Swift}/XRT in X-rays (0.2$-$10\,keV).
}
\label{fig:flimit}
\end{figure*}

Figure~\ref{fig:flimit} plots the frequency distribution of the triple disk flux density for different binary masses (panel a), distances (panel b), and CBD's outer radii (panel c) superimposed on the observational flux limits. The reference model is the flux density with $M=100M_\odot$, $D=10\,{\rm Mpc}$, and $r_{\rm out}/r_{\rm in}=1000$.

Panel (a) shows that a triple disk of stellar to intermediate-mass black holes is detectable in the IR, optical, and X-ray at a distance of 10 Mpc from Earth. As expected, the spectral magnitude is higher with increasing binary mass. 
Panel (b) indicates that the reference model is too dark to be observed in any waveband at 100 Mpc, but the $10^3 M_\odot$ case is detectable even at 100 Mpc.

Panel (c) demonstrates that the brightness of the low-frequency peaks varies with the outer radius of the CBD, while that of the high-frequency peaks does not at all. This makes sense because the high-frequency peak comes from the minidisks and is therefore independent of the outer radius of the CBD. In particular, the outer radius as small as $10^2\,r_\mathrm{in}$ makes the low-frequency peak weaker than the high-frequency peak. Nevertheless, even this smallest CBD case is sufficiently observable with JWST/NIRCam and {\it Swift}/XRT.

Let us consider the brief strategy of Target-of-Opportunity observations to test our model: If {\it Swift}/XRT finds a bright flare in X-rays (within 100 Mpc if the distance to the source is available), we prompt NIR telescopes to look into that region.
If a point source is found, Subaru and JWST will be pointed at it to examine the detailed frequency dependence of the flux density. This will allow us to estimate the power-law index of the flux density in the NIR band and compare it with our model.

%
\section{Discussion}
\label{eq:diss}
%
Considering the conservation of radiative flux in the vertical direction of the CBD, the vertical flux at the disk mid-plane is equivalent to the flux at the surface, giving \citep{kato_black-hole_2008} 
\begin{eqnarray}
T
=
\frac{2}{3^{1/4}}\frac{1}{\tau^{1/4}}
T_{\rm c}
\label{eq:tc}
\end{eqnarray}
where, according to the standard disk model, the optical depth is given by
\begin{eqnarray}
  \tau
  &&
  =\frac{\kappa\Sigma}{2}
  \nonumber \\
  &&
  \sim
  4.6
  \left(\frac{M}{100\,M_\odot}\right)^{1/2}
  \left(\frac{\kappa}{0.4\,{\rm cm^2\,g^{-1}}}\right)
  \left(\frac{\alpha_{\rm SS}}{0.1}\right)^{-1}
  \left(\frac{H/r}{0.01}\right)^{-1}
  \left(\frac{\dot{m}}{1.0}\right)^{-1}
  \left(\frac{r}{r_{\rm b}}\right)^{-1/2}
\end{eqnarray}
with the surface density $\Sigma=\dot{M}/(3\pi \nu)$ and the disk viscosity $\nu=(2/3)\alpha_{\rm SS}c_{\rm s}H$ with the Shakura-Sunayev viscosity parameter $\alpha_{\rm SS}$ (see also equation~\ref{eq:alphaviscosity}). Here, we used equation~(\ref{eq:hydrostaticeq}) for the derivation. Assuming that the CBD opacity source is electron scattering, i.e., $\kappa=0.4\,{\rm cm^2\,g^{-1}}$ the optical depth is of the order of unity for $r \ge r_{\rm b}$, resulting in $T\approx T_{\rm c}$ in the irradiation heating dominated region from the equation~(\ref{eq:tc}). In this case, our assumption of $T\approx{T_{\rm c}}$ is justified. However, as the CBD mid-plane temperature drops to several thousand degrees Kelvin, the free-free absorption comes into play, and the difference between the disk mid-plane temperature and the surface temperature is expected to become significant. At lower temperatures, the bound-free absorption is also an important source of opacity. In other words, the spectra of the irradiated part of the CBD will vary with these opacities. Since, for example, the opacity due to the free-free absorption has a complicated dependence on the density and temperature as $\kappa\propto\rho{T}_{\rm c}^{-3.5}$, the differential equation for the disk aspect ratio becomes too complicated to solve analytically, unlike equation~(\ref{eq:dydx1}). In a forthcoming paper, we will numerically study the CBD spectra by taking these opacities into account.

%


Next, we discuss the effect of the CBD inner edge radius on the spectrum. Note that the inner edge radius is proportional to $a_0$ and $C_{\rm gap}$. As the inner-edge radius increases, the inner-edge temperature will be lower, shifting the high-frequency peak of the CBD toward the lower frequencies. In contrast, the low frequency peak will be relatively prominent. If the inner-edge radius is smaller, the inner-edge temperature will be higher. This would make the high frequency peak larger. 

The timescale for the binary orbital decay due to gravitational wave radiation is given by \citep{Peters64} 
\begin{eqnarray}
t_{\rm gw}
&&
=
\frac{5}{8}
\left(\frac{a}{r_{\rm S}}\right)^4
\left(\frac{r_{\rm S}}{c}\right)
g(q)
\nonumber \\
&&
\sim\,7.8\times10\,{\rm yr}
\left(\frac{M}{100\,M_\odot}\right)
\left(\frac{a_0}{1000}\right)^{4}
\left(\frac{g(q)}{4}\right),
\nonumber \\
\label{eq:tgw}
\end{eqnarray}
where $g(q)=(1+q)^2/q$ and $g(q=1)=4$.
This suggests that the binary with $a=1000r_{\rm S}$ is on the way to merging with radiating gravitational waves. Now consider a situation where the CBD and binary are dynamically coupled; if the CBD viscosity time estimated at the inner edge of the CBD is longer than the coalescence time, the CBD and binary will move rapidly toward coalescence, leaving the CBD to decouple. For our fiducial model, the semi-major axis of the decoupled binary is estimated to be about $100r_{\rm S}$. If the difference between the pre- and post-decoupled spectra is pronounced, we can distinguish between binaries dynamically with and without CBDs. This is also a future topic, including scaling the black hole mass to the supermassive black holes (SMBHs).


%
\section{Conclusions}
\label{sec:conclusion}
%

We have studied the effect of irradiation from the two minidisks on the circumbinary disk (CBD). We have derived the irradiation heating rate on the CBD surface and then, by considering the energy balance equation, a first-order differential equation for the radial profile of the disk aspect ratio. Assuming the hydrostatic equilibrium of the CBD in the vertical direction and adopting a consistent outer boundary condition, we have solved the differential equation analytically and numerically. Using these solutions, we have calculated the CBD spectra and studied their dependence on the black hole mass, the binary mass ratio, and the CBD outer radius. We also computed the combined disk spectrum, which is the sum of the CBD spectrum and two minidisk spectra, the so-called triple disk spectra. Our conclusions are summarized as follows:
\begin{enumerate}
\item
The analytically calculated spectra are in good agreement with the numerical spectra.

\item
The CBD spectrum is almost independent of the binary mass ratio.

\item
Triple disk spectra show double peaks. The high frequency peak arises from the two minidisks, while the low frequency peak from the irradiated CBD. Note that the latter peak does not appear when the CBD is not irradiated. For the stellar-mass to intermediate-mass range of binary black holes, the high frequency peak appears in the soft X-ray band, while the low frequency peak appears in the IR band.

\item
The observed flux increases with frequency as $\nu^{1/3}$ towards the high-frequency peak, while it decreases with frequency away from the low-frequency peak as $\nu^{-5/3}$. The latter feature can be tested with near-IR observations with Subaru and JWST.
\end{enumerate}
%
\section*{Acknowledgments}
%
\begin{acknowledgments}
Y.L. and K.H. have been supported by the Basic Science Research Program through the National Research Foundation of Korea (NRF) funded by the Ministry of Education (2016R1A5A1013277 to K.H. and 2020R1A2C1007219 to K.H. and Y.L.). This research was also supported in part by grant no. NSF PHY-2309135 to the Kavli Institute for Theoretical Physics (KITP), and also supported by Grant-inAid for Scientific Research from the MEXT/JSPS of Japan, JP21K03619 to A.T.O.
\end{acknowledgments}

\appendix

%
\section{Irradiated fluxes from two minidisks}
\label{app:A}
\begin{figure*}[htbp]
    \centering
    \includegraphics[scale=0.28]{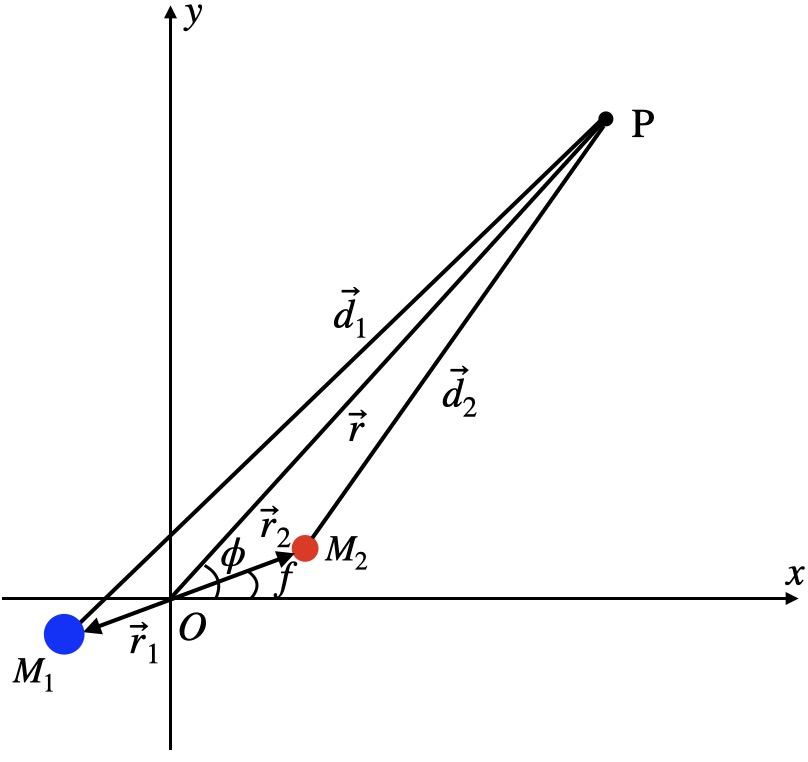}
    \includegraphics[scale=0.28]{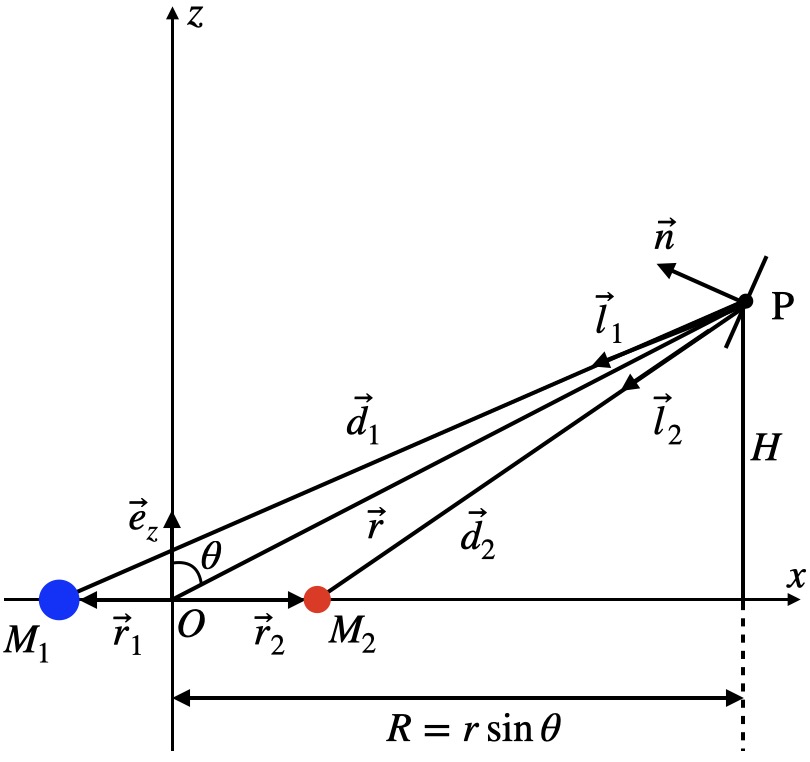}
    \caption{
A schematic view of a binary black hole and its surrounding CBD. The left panel illustrates a binary black hole moving in the xy plane and the CBD surrounding it in the face-on view, while the right panel does the same as the left panel but in the xz plane, i.e. from the edge-on view. The blue and red filled circles denote the primary and secondary black holes with their masses $M_1$ and $M_2$, respectively. We use polar coordinates for the vectors and the origin is the center of mass of the binary.  In both panels, $\vec r_1$ and $\vec r_2$ are the position vectors from the origin to each BH, and $\vec r$ is a position vector from the origin to point $P$ on the CBD surface. Here $\vec d_1=\vec r-\vec r_1$, $\vec d_2=\vec r-\vec r_2$, $\phi$ is an azimuthal angle, and $f$ is the true anomaly of the binary motion. 
In the right panel, $H$ is the scale height of the CBD, $\vec R$ is the position vector from the origin to the point $p$ projected on the $x$ axis, and $\theta$ is a tilt angle. $\vec n$ is the unit vector of the vertical direction of the CBD surface, $\vec l_1$ is the unit vector of $\vec d_1$, and $\vec e_z$ is the unit vector along the $z$ axis. 
    }
    \label{fig:CBDgeom}
\end{figure*}

%
We calculate the irradiation flux directed toward the CBD from the accretion disks (i.e., minidisks) around the primary and secondary black holes. We assume that the two sources are point sources and orbit in a circular binary. This is because they are close to the inner edge of each disk 
and much smaller than the orbital semi-major axis of the binary black hole. The flux from the two black holes to the CBD surface is expected to vary with the orbital motion of the binary as their distance varies. To properly calculate the flux, the positional relationship between the binary and the CBD surface is shown in Figure~\ref{fig:CBDgeom}.

The vectors of Figure~\ref{fig:CBDgeom} are expressed in polar coordinates as follows:
\begin{eqnarray}
    \vec r
    &&
    =
    r\sin\theta\cos\phi \, \hat x+r\sin\theta\sin\phi \, \hat y+r\cos\theta \, \hat z
    \label{eq:rvec}
    \\
    \vec r_1
    &&
    =r_1\cos f\, \hat x+r_1\sin f \, \hat y
    \label{eq:r1vec}
    \\
    \vec r_2
    &&
    =r_2\cos (\pi+f)\,\hat x+r_2\sin(\pi+f) \hat y=-r_2\cos f\,\hat x-r_2\sin f\, \hat y 
    \label{eq:r2vec}
    \\
    \vec d_1
    &&
    =(r\sin\theta\cos\phi-r_1\cos f)\,\hat x+(r\sin\theta\sin\phi-r_1\sin f)\,\hat y+r\cos\theta\,\hat z 
    \label{eq:d1vec}
    \\
    \vec d_2
    &&
    =(r\sin\theta\cos\phi+r_2\cos f)\,\hat x+(r\sin\theta\sin\phi+r_2\sin f)\,\hat y+r\cos\theta\,\hat z 
    \label{eq:d2vec}
    \\
    \vec n
    &&
    =\left[1+\left(\frac{dH}{dR}\right)^2\right]^{-1/2}
    \left(-\frac{dH}{dR}\cos\phi,\, -\frac{dH}{dR}\sin\phi,\, 1 \right)
    \label{eq:nvec}
\end{eqnarray}
with the assumption $H\ll R$ and $f$ is the true anomaly of the binary orbit, where
\begin{eqnarray}
     r_1&=&\frac{q}{1+q}a 
     \label{eq:r1}
     \\
     r_2&=&\frac{1}{1+q}a
     \label{eq:r2}
\end{eqnarray}
with the mass ratio $q=M_1/M_2$ and the semi-major axis $a$.

The unit vectors of $\vec{d}_1$ and $\vec{d}_2$ are given by 
\begin{equation}
    \vec l_1
    =\frac{-\vec d_1}{|\vec d_1|},
    \hspace{0.5cm}
    \vec l_2
    =
    \frac{-\vec d_2}{|\vec d_2|}, 
    \label{eq:l1l2}
\end{equation}
respectively, where
\begin{eqnarray}
    \left|\vec{d}_1\right| &=& \left[ r^2 + r_1^2 - 2rr_1 \sin\theta \cos(\phi - f) \right]^{1/2} 
    \label{eq:d1}
    \\
    \left|\vec{d}_2\right| &=& \left[ r^2 + r_2^2 - 2rr_2 \sin\theta \cos(\phi - f) \right]^{1/2}.
    \label{eq:d2}
\end{eqnarray}

%
\subsection{Irradiated fluxes}
%
From the geometrical relation of Figure~\ref{fig:CBDgeom}, the irradiated flux on the CBD surface from each BH is given by 
\begin{equation}\label{eq:irflux2}
    F_{\rm irr,i}=\frac{L_i}{4\pi D^2}(\vec l_i\cdot \vec n)(\vec n \cdot \vec e_z),
\end{equation} 
where
$i=1$ and $i=2$ denote the primary and secondary black hole, respectively, $D$ is the distance from Earth. Here, the bolometric luminosity for each black hole $L_i$ is given by
\begin{eqnarray}
    L_i\equiv \frac{GM_i\dot M_i}{r_{{\rm ISCO},i}},
    \label{eq:lumii}
\end{eqnarray}
where $\dot M_i$, $M_{i}$, and $r_{{\rm ISCO},i}=6GM_i/c^2$ are the mass accretion rate, black hole mass, and ISCO radius of each black hole.

From equations~(\ref{eq:l1l2}) and (\ref{eq:nvec}), we get
\begin{eqnarray}
    \vec{l}_1 \cdot \vec{n} 
    &=& \left[1 + \left( \frac{dH}{dR} \right)^2 \right]^{-1/2} d_1^{-1} 
    \left( r \frac{dH}{dr} - r_1 \frac{dH}{dr} \frac{\cos(\phi - f)}{\sin\theta} - r \cos\theta \right) 
    \label{eq:l1n}
    \nonumber \\
    \vec{l}_2 \cdot \vec{n} 
    &=& \left[1 + \left( \frac{dH}{dR} \right)^2 \right]^{-1/2} d_2^{-1} 
    \left( r \frac{dH}{dr} + r_2 \frac{dH}{dr} \frac{\cos(\phi - f)}{\sin\theta} - r \cos\theta \right)
    \label{eq:l2n}
    \\
    \vec{n} \cdot \vec{e}_z &=& \left[1 + \left( \frac{dH}{dR} \right)^2 \right]^{-1/2}, 
    \label{eq:nez}
    \\
    \frac{dH}{dR}
    &=&\frac{dH}{dr}\frac{1}{\sin\theta}.
    \label{eq:dhdrlarge}
\end{eqnarray}
Substituting equations~(\ref{eq:l1n})-(\ref{eq:nez}) into the equation (\ref{eq:irflux2}) gives
\begin{eqnarray}
    F_{\rm irr,1} 
    &=& F_1(\vec l_1 \cdot \vec n)(\vec n \cdot \vec e_z) 
    \nonumber\\
    &=& \frac{L_1}{4\pi d_1^3}
    \left[1+\left(\frac{dH}{dR}\right)^2\right]^{-1}
    \left(r\frac{dH}{dr}-r_1\frac{dH}{dr}\frac{\cos(\phi-f)}{\sin\theta}-r\cos\theta\right)
    \nonumber \\
    &=&
    \frac{L_1}{4\pi r^2}\left[1-\left(\frac{dH}{dr}\right)^2\right]
    \left[\frac{dH}{dr}-\frac{r_1}{r}
    \frac{dH}{dr}\frac{\cos(\phi-f)}{\sin\theta}
    -\frac{H}{r}\right. \nonumber \\
    &&\left. +3\frac{r_1}{r}\frac{dH}{dr}\sin\theta\cos(\phi-f)-3\left(\frac{r_1}{r}\right)^2\frac{dH}{dr}\cos^2(\phi-f)-\frac{3}{2}\frac{r_1}{r}\sin2\theta\cos(\phi-f)\right.
    \nonumber \\ 
    &&\left. +\frac{3}{2}\left(\frac{r_1}{r}\right)^2(5\cos^2(\phi-f)-1)\left(\frac{dH}{dr}-\frac{r_1}{r}\frac{dH}{dr}\frac{\cos(\phi-f)}{\sin\theta}-\frac{H}{r}\right)\right],
    \label{eq:firr1}
\\
    F_{\rm irr,2} &=& F_2(\vec l_2 \cdot \vec n)(\vec n \cdot \vec e_z) \nonumber \\
    &=& \frac{L_2}{4\pi d_2^3} \left[1+\left(\frac{dH}{dR}\right)^2\right]^{-1} 
    \left(r\frac{dH}{dr}+r_2\frac{dH}{dr}\frac{\cos(\phi-f)}{\sin\theta}-r\cos\theta\right)
    \nonumber \\
    &=&\frac{L_2}{4\pi r^2}\left[1-\left(\frac{dH}{dr}\right)^2\right]\left[ \frac{dH}{dr}+\frac{r_2}{r}\frac{dH}{dr}\frac{\cos(\phi-f)}{\sin\theta}-\frac{H}{r} \right.\nonumber \\
    &&\left. -3\frac{r_2}{r}\frac{dH}{dr}\sin\theta\cos(\phi-f)-3\left(\frac{r_2}{r}\right)^2\frac{dH}{dr}\cos^2(\phi-f)+\frac{3}{2}\frac{r_2}{r}\sin2\theta\cos(\phi-f) \right.
    \nonumber \\
    &&\left. +\frac{3}{2}\left(\frac{r_2}{r}\right)^2 (5\cos^2(\phi-f)-1) \left(\frac{dH}{dr}+\frac{r_2}{r}\frac{dH}{dr}\frac{\cos(\phi-f)}{\sin\theta}-\frac{H}{r}\right)\right],
    \label{eq:firr2}
\end{eqnarray}
where the term $d_i$ is approximately obtained in a Taylor series of $r_i/r$ as
\begin{eqnarray}
\label{eq:d1cubic}
    \frac{1}{d_1^3}
    &=&[r^2+r_1^2-2rr_1\sin\theta\cos(\phi-f)]^{-3/2}\nonumber \\
    &\approx& r^{-3} \left[1+3\frac{r_1}{r}\sin\theta\cos(\phi-f)
    +\frac{3}{2}\left(\frac{r_1}{r}\right)^2\left(5\sin^2\theta\cos^2(\phi-f)-1\right)\right], \\
\label{eq:d2cubic}
    \frac{1}{d_2^3}
    &=&[r^2+r_2^2+2rr_2\sin\theta\cos(\phi-f)]^{-3/2}\nonumber \\
    &\approx& r^{-3}\left[1-3\frac{r_2}{r}\sin\theta\cos(\phi-f)
    +\frac{3}{2}\left(\frac{r_2}{r}\right)^2\left(5\sin^2\theta\cos^2(\phi-f)-1\right)\right],
\end{eqnarray}
repectively.

%
\section{Irradiation heating rates}
%
The irradiated heating rate is given by 
\begin{equation}
    Q_{{\rm irr},i}=2A_i F_{{\rm irr},i},
    \label{eq:qirr}
\end{equation}
where the factor $2$ indicates that the irradiated flux heats the two sides of the CBD surface, and $A_i$ represents the ratio of re-emitted to incident photons. 
The CBD viscous timescale at $r_{\rm b}$ and binary orbital period are evaluated as
\begin{eqnarray}
    \tau_{\rm vis}
    &&
    =\frac{r^2}{\nu}
    \approx
    \frac{3}{2}
    \frac{r^2\Omega}{\alpha_{\rm SS}\,c_{\rm s}^2}
    \nonumber \\
    &&
    \sim
    1.9
    \times
    10^9
    \,
    {\rm s}
    \,
    \left(\frac{r}{r_{\rm b}}\right)^{1/2}
    \left(\frac{\alpha_{\rm SS}}{0.1}\right)^{-1}
    \left(\frac{M}{100\,M_\odot}\right)^{25/18}
    \left(\frac{\dot{m}}{1.0}\right)^{-7/18}
    \left(\frac{A}{0.1}\right)^{-10/9},
    \label{eq:vistime}
    \\
    P_{\rm orb}
    &&
    =\frac{2\pi}{\Omega}
    \nonumber \\
    &&
    \sim
    2.8
    \times
    10^2
    \,
    {\rm s}
    \,
    \left(\frac{a_0}{1000}\right)^{3/2}
    \left(\frac{M}{100\,M_\odot}\right),
    \label{eq:bop}
    \end{eqnarray}
    where the disk viscosity is given by
    \begin{eqnarray}
    \nu
    &&
   = \frac{2}{3}\alpha_{\rm SS}c_{\rm s}H.
   \label{eq:alphaviscosity}
\end{eqnarray}
with the Shacla-Sunayev viscosity parameter $\alpha_{\rm SS}$ \citep{1973A&A....24..337S}.
Since the viscous timescale of the CBD is much longer than the binary orbital period, and also the binary is in a circular orbit, the effect of the binary motion on the irradiation heating to the CBD can be considered quasi-stationary. Thus, the irradiation heating rate is approximately equal to the azimuthally- and orbit-averaged irradiation heating rate:

\begin{equation}
    \left<Q_{\rm irr,i}\right>=\frac{1}{2\pi}\int^{2\pi}_0 Q_{\rm irr,i}\,d\phi',
    \label{eq:aveqirr}
\end{equation}
where $(\phi'=\phi-f)$. 
From equations (\ref{eq:firr1}), (\ref{eq:firr2}), (\ref{eq:qirr}), and (\ref{eq:aveqirr}), we get 
$\left<Q_{\rm irr,1}\right>$ and $\left<Q_{\rm irr,2}\right>$ as
\begin{eqnarray}
\label{eq:qirr1}
    \left<Q_{\rm irr,1}\right> 
    &=& 
    \frac{A_1L_1}{2\pi r^2} 
    \left[1-\left(\frac{dH}{dr}\right)^2\right]
    \left[
    \frac{dH}{dr}-\frac{H}{r}-\frac{3}{2}
    \left(\frac{r_1}{r}\right)^2\frac{dH}{dr}-\frac{3}{2}\left(\frac{r_1}{r}\right)^2
    \left(\frac{dH}{dr}-\frac{H}{r}\right)
    \right. \nonumber \\
    && \hspace{1cm}
    \left.
    +\frac{15}{4}\left(\frac{r_1}{r}\right)^2\left(\frac{dH}{dr}-\frac{H}{r}\right)
    \right] 
    \nonumber\\
    &\approx& 
    \frac{A_1L_1}{2\pi r} 
    \Biggr[
    \frac{d}{dr}\left(\frac{H}{r}\right)+
    \frac{3}{2}\left(\frac{r_1}{r}\right)^2\left[\frac{3}{2}\frac{d}{dr}\left(\frac{H}{r}\right)-\frac{1}{r}\frac{dH}{dr}\right]
    \Biggr]
    \nonumber \\
    &=& \frac{A_1L_1}{2\pi r} 
     \Biggr[
     \frac{d}{dr}\left(\frac{H}{r}\right)
     -
    \beta_1\left(\frac{r_{\rm in}}{r}\right)^2
    \left[
    \frac{H}{r^2}
    -
    \frac{1}{2}\frac{d}{dr}\left(\frac{H}{r}\right)
    \right]
    \Biggr],
\\
    \left<Q_{\rm irr,2}\right>
    &=&
    \frac{A_2L_2}{2\pi r^2} \left[1-\left(\frac{dH}{dr}\right)^2\right]\left[\frac{dH}{dr}-\frac{H}{r}-\frac{3}{2}\left(\frac{r_2}{r}\right)^2\frac{dH}{dr}-\frac{3}{2}\left(\frac{r_2}{r}\right)^2\left(\frac{dH}{dr}-\frac{H}{r}\right)\right. \nonumber \\
    &&\hspace{1cm}\left.+\frac{15}{4}\left(\frac{r_2}{r}\right)^2\left(\frac{dH}{dr}-\frac{H}{r}\right)
    \right] 
    \nonumber\\
    &\approx& 
    \frac{A_2L_2}{2\pi r} 
    \Biggr[
    \frac{d}{dr}\left(\frac{H}{r}\right)+
    \frac{3}{2}\left(\frac{r_2}{r}\right)^2
    \left[
    \frac{3}{2}\frac{d}{dr}\left(\frac{H}{r}\right)-\frac{1}{r}\frac{dH}{dr}
    \right]
    \Biggr]
    \nonumber \\
    &=& 
    \frac{A_2L_2}{2\pi r} 
     \Biggr[
     \frac{d}{dr}\left(\frac{H}{r}\right)
     -
    \beta_2\left(\frac{r_{\rm in}}{r}\right)^2
    \left[
    \frac{H}{r^2}
    -
    \frac{1}{2}\frac{d}{dr}\left(\frac{H}{r}\right)
    \right]
    \Biggr],
    \label{eq:qirr2}
\end{eqnarray}
where $\beta_1$ and $\beta_2$ are given by
\begin{eqnarray}
    \beta_1=\frac{3}{2}\left(\frac{r_1}{r_{\rm in}}\right)^2
   =\frac{3}{2}\frac{1}{C_{\rm gap}^2}\frac{q^2}{(1+q)^2}
    , \,\,
    \beta_2=\frac{3}{2}\left(\frac{r_2}{r_{\rm in}}\right)^2
      =\frac{3}{2}\frac{1}{C_{\rm gap}^2}\frac{1}{(1+q)^2},
    \label{eq:beta-param}
\end{eqnarray}
respectively, and $\beta_1$ and $\beta_2$ are less than unity for $q\le1.0$. 

From equations (\ref{eq:qirr1}) and (\ref{eq:qirr2}), we obtain the total irradiated heating as
\begin{eqnarray}
    Q_{\rm irr}
    &=&
    \frac{A_1L_1}{2\pi r} 
     \Biggr[
     \frac{d}{dr}\left(\frac{H}{r}\right)
     -
    \beta_1\left(\frac{r_{\rm in}}{r}\right)^2
    \left[
    \frac{H}{r^2}
    -
    \frac{1}{2}\frac{d}{dr}\left(\frac{H}{r}\right)
    \right]
    \Biggr]
    +
    \frac{A_2L_2}{2\pi r} 
     \Biggr[
     \frac{d}{dr}\left(\frac{H}{r}\right)
     -
    \beta_2\left(\frac{r_{\rm in}}{r}\right)^2
    \left[
    \frac{H}{r^2}
    -
    \frac{1}{2}\frac{d}{dr}\left(\frac{H}{r}\right)
    \right]
    \Biggr].
    \nonumber \\
    \label{eq:qirrtot}
\end{eqnarray}
%
%
\section{Analytical solutions for $\beta\neq0$}
\label{app:anasol}
%
Equation~(\ref{eq:dydx1}) can be integrated analytically. By separating the variables, we obtain  
\begin{eqnarray}
\int_{Y_{\rm b}}^{Y}
    \frac{1}{(\alpha Y^8+\beta{Y})}
    dY
    =
    \int_{\xi_{\rm b}}^{\xi}
    \frac{1}{(\xi^3+(\beta/2)\xi)}
    d\xi,
    \nonumber
\end{eqnarray}
where $Y_{\rm b}=Y(\xi_{\rm b})$ and $\xi_{\rm b}$ is the normalized boundary radius, which can be taken arbitrarily in this formulation. Both sides are integrated as
\begin{eqnarray}
    \log\frac{Y}{(\alpha{Y^7}+\beta)^{1/7}}
    -
    \log\frac{Y_{\rm b}}{(\alpha{Y_{\rm b}^7}+\beta)^{1/7}}
    =
    \log\frac{\xi^2}{2\xi^2+\beta}
    -
     \log\frac{\xi_{\rm b}^2}{2\xi_{\rm b}^2+\beta}.
     \nonumber 
\end{eqnarray}
This equation gives the general solution as
\begin{eqnarray}
   Y=\beta^{1/7}\left(\frac{J\,\xi^{2}}
    {\beta+2{\xi^2}}\right)
    \left[ 1-\alpha\left(\frac{J\,\xi^{2}}
    {\beta + 2{\xi^2}}\right)^7\right]^{-1/7},
    \label{eq:yanalsol}
\end{eqnarray}
where the integral constant $J$ is given by
\begin{eqnarray}
    J=\frac{2\xi_{\rm b}^2+\beta}{\xi_{\rm b}^2}\frac{Y_{\rm b}}{(\alpha{Y_{\rm b}^7}+\beta)^{1/7}}.
    \nonumber 
\end{eqnarray}
It follows from equation (\ref{eq:yanalsol}) that 
the global $(1\le{\xi}\le\infty)$ solutions exist only if the following condition is satisfied:
\begin{eqnarray}
    \frac{\alpha^{1/7}}{2}J\le1.
    \nonumber 
\end{eqnarray} 
Here, substituting $J=2/\alpha^{1/7}$ into equation~(\ref{eq:yanalsol}) yields the following particular solution:
\begin{eqnarray}
Y=\left(\frac{\beta}{\alpha}\right)^{1/7}
\left[1+\frac{1}{2}\frac{\beta}{\xi^2}\right]^{-1}
\left[
1-\left(1+\frac{\beta}{2\xi^2}\right)^{-7}
\right]^{-1}.
\label{eq:yanalsol2}
\end{eqnarray}
Taking $\beta=0$ at the limit of $\beta/\xi^2 \to 0$, equation~(\ref{eq:yanalsol2}) consistently connects to the power-law solution of equation~(\ref{eq:appsol2}).

%
\bibliography{yw}{}

\begin{thebibliography}{}
\expandafter\ifx\csname natexlab\endcsname\relax\def\natexlab#1{#1}\fi
\providecommand{\url}[1]{\href{#1}{#1}}
\providecommand{\dodoi}[1]{doi:~\href{http://doi.org/#1}{\nolinkurl{#1}}}
\providecommand{\doeprint}[1]{\href{http://ascl.net/#1}{\nolinkurl{http://ascl.net/#1}}}
\providecommand{\doarXiv}[1]{\href{https://arxiv.org/abs/#1}{\nolinkurl{https://arxiv.org/abs/#1}}}

\bibitem[{{Abbott} {et~al.}(2016){Abbott}, {Abbott}, {Abbott}, {Abernathy},
  {Acernese}, {Ackley}, {Adams}, {Adams}, {Addesso}, {Adhikari}, {Adya},
  {Affeldt}, {Agathos}, {Agatsuma}, {Aggarwal}, {Aguiar}, {Aiello}, {Ain},
  {Ajith}, {Allen}, {Allocca}, {Altin}, {Anderson}, {Anderson}, {Arai},
  {Arain}, {Araya}, {Arceneaux}, {Areeda}, {Arnaud}, {Arun}, {Ascenzi},
  {Ashton}, {Ast}, {Aston}, {Astone}, {Aufmuth}, {Aulbert}, {Babak}, {Bacon},
  {Bader}, {Baker}, {Baldaccini}, {Ballardin}, {Ballmer}, {Barayoga},
  {Barclay}, {Barish}, {Barker}, {Barone}, {Barr}, {Barsotti}, {Barsuglia},
  {Barta}, {Bartlett}, {Barton}, {Bartos}, {Bassiri}, {Basti}, {Batch},
  {Baune}, {Bavigadda}, {Bazzan}, {Behnke}, {Bejger}, {Belczynski}, {Bell},
  {Bell}, {Berger}, {Bergman}, {Bergmann}, {Berry}, {Bersanetti}, {Bertolini},
  {Betzwieser}, {Bhagwat}, {Bhandare}, {Bilenko}, {Billingsley}, {Birch},
  {Birney}, {Birnholtz}, {Biscans}, {Bisht}, {Bitossi}, {Biwer}, {Bizouard},
  {Blackburn}, {Blair}, {Blair}, {Blair}, {Bloemen}, {Bock}, {Bodiya}, {Boer},
  {Bogaert}, {Bogan}, {Bohe}, {Bojtos}, {Bond}, {Bondu}, {Bonnand}, {Boom},
  {Bork}, {Boschi}, {Bose}, {Bouffanais}, {Bozzi}, {Bradaschia}, {Brady},
  {Braginsky}, {Branchesi}, {Brau}, {Briant}, {Brillet}, {Brinkmann},
  {Brisson}, {Brockill}, {Brooks}, {Brown}, {Brown}, {Brown}, {Buchanan},
  {Buikema}, {Bulik}, {Bulten}, {Buonanno}, {Buskulic}, {Buy}, {Byer},
  {Cabero}, {Cadonati}, {Cagnoli}, {Cahillane}, {Bustillo}, {Callister},
  {Calloni}, {Camp}, {Cannon}, {Cao}, {Capano}, {Capocasa}, {Carbognani},
  {Caride}, {Casanueva Diaz}, {Casentini}, {Caudill}, {Cavagli{\`a}},
  {Cavalier}, {Cavalieri}, {Cella}, {Cepeda}, {Baiardi}, {Cerretani},
  {Cesarini}, {Chakraborty}, {Chalermsongsak}, {Chamberlin}, {Chan}, {Chao},
  {Charlton}, {Chassande-Mottin}, {Chen}, {Chen}, {Cheng}, {Chincarini},
  {Chiummo}, {Cho}, {Cho}, {Chow}, {Christensen}, {Chu}, {Chua}, {Chung},
  {Ciani}, {Clara}, {Clark}, {Cleva}, {Coccia}, {Cohadon}, {Colla}, {Collette},
  {Cominsky}, {Constancio}, {Conte}, {Conti}, {Cook}, {Corbitt}, {Cornish},
  {Corsi}, {Cortese}, {Costa}, {Coughlin}, {Coughlin}, {Coulon}, {Countryman},
  {Couvares}, {Cowan}, {Coward}, {Cowart}, {Coyne}, {Coyne}, {Craig},
  {Creighton}, {Creighton}, {Cripe}, {Crowder}, {Cruise}, {Cumming},
  {Cunningham}, {Cuoco}, {Dal Canton}, {Danilishin}, {D'Antonio}, {Danzmann},
  {Darman}, {Da Silva Costa}, {Dattilo}, {Dave}, {Daveloza}, {Davier},
  {Davies}, {Daw}, {Day}, {De}, {DeBra}, {Debreczeni}, {Degallaix}, {De
  Laurentis}, {Del{\'e}glise}, {Del Pozzo}, {Denker}, {Dent}, {Dereli},
  {Dergachev}, {DeRosa}, {De Rosa}, {DeSalvo}, {Dhurandhar}, {D{\'\i}az}, {Di
  Fiore}, {Di Giovanni}, {Di Lieto}, {Di Pace}, {Di Palma}, {Di Virgilio},
  {Dojcinoski}, {Dolique}, {Donovan}, {Dooley}, {Doravari}, {Douglas},
  {Downes}, {Drago}, {Drever}, {Driggers}, {Du}, {Ducrot}, {Dwyer}, {Edo},
  {Edwards}, {Effler}, {Eggenstein}, {Ehrens}, {Eichholz}, {Eikenberry},
  {Engels}, {Essick}, {Etzel}, {Evans}, {Evans}, {Everett}, {Factourovich},
  {Fafone}, {Fair}, {Fairhurst}, {Fan}, {Fang}, {Farinon}, {Farr}, {Farr},
  {Favata}, {Fays}, {Fehrmann}, {Fejer}, {Feldbaum}, {Ferrante}, {Ferreira},
  {Ferrini}, {Fidecaro}, {Finn}, {Fiori}, {Fiorucci}, {Fisher}, {Flaminio},
  {Fletcher}, {Fong}, {Fournier}, {Franco}, {Frasca}, {Frasconi}, {Frede},
  {Frei}, {Freise}, {Frey}, {Frey}, {Fricke}, {Fritschel}, {Frolov}, {Fulda},
  {Fyffe}, {Gabbard}, {Gair}, {Gammaitoni}, {Gaonkar}, {Garufi}, {Gatto},
  {Gaur}, {Gehrels}, {Gemme}, {Gendre}, {Genin}, {Gennai}, {George}, {Gergely},
  {Germain}, {Ghosh}, {Ghosh}, {Ghosh}, {Giaime}, {Giardina}, {Giazotto},
  {Gill}, {Glaefke}, {Gleason}, {Goetz}, {Goetz}, {Gondan}, {Gonz{\'a}lez},
  {Castro}, {Gopakumar}, {Gordon}, {Gorodetsky}, {Gossan}, {Gosselin},
  {Gouaty}, {Graef}, {Graff}, {Granata}, {Grant}, {Gras}, {Gray}, {Greco},
  {Green}, {Greenhalgh}, {Groot}, {Grote}, {Grunewald}, {Guidi}, {Guo},
  {Gupta}, {Gupta}, {Gushwa}, {Gustafson}, {Gustafson}, {Hacker}, {Hall},
  {Hall}, {Hammond}, {Haney}, {Hanke}, {Hanks}, {Hanna}, {Hannam}, {Hanson},
  {Hardwick}, {Harms}, {Harry}, {Harry}, {Hart}, {Hartman}, {Haster},
  {Haughian}, {Healy}, {Heefner}, {Heidmann}, {Heintze}, {Heinzel}, {Heitmann},
  {Hello}, {Hemming}, {Hendry}, {Heng}, {Hennig}, {Heptonstall}, {Heurs},
  {Hild}, {Hoak}, {Hodge}, {Hofman}, {Hollitt}, {Holt}, {Holz}, {Hopkins},
  {Hosken}, {Hough}, {Houston}, {Howell}, {Hu}, {Huang}, {Huerta}, {Huet},
  {Hughey}, {Husa}, {Huttner}, {Huynh-Dinh}, {Idrisy}, {Indik}, {Ingram},
  {Inta}, {Isa}, {Isac}, {Isi}, {Islas}, {Isogai}, {Iyer}, {Izumi}, {Jacobson},
  {Jacqmin}, {Jang}, {Jani}, {Jaranowski}, {Jawahar}, {Jim{\'e}nez-Forteza},
  {Johnson}, {Johnson-McDaniel}, {Jones}, {Jones}, {Jonker}, {Ju}, {Haris},
  {Kalaghatgi}, {Kalogera}, {Kandhasamy}, {Kang}, {Kanner}, {Karki},
  {Kasprzack}, {Katsavounidis}, {Katzman}, {Kaufer}, {Kaur}, {Kawabe},
  {Kawazoe}, {K{\'e}f{\'e}lian}, {Kehl}, {Keitel}, {Kelley}, {Kells},
  {Kennedy}, {Keppel}, {Key}, {Khalaidovski}, {Khalili}, {Khan}, {Khan},
  {Khan}, {Khazanov}, {Kijbunchoo}, {Kim}, {Kim}, {Kim}, {Kim}, {Kim}, {Kim},
  {King}, {King}, {Kinzel}, {Kissel}, {Kleybolte}, {Klimenko}, {Koehlenbeck},
  {Kokeyama}, {Koley}, {Kondrashov}, {Kontos}, {Koranda}, {Korobko}, {Korth},
  {Kowalska}, {Kozak}, {Kringel}, {Krishnan}, {Kr{\'o}lak}, {Krueger}, {Kuehn},
  {Kumar}, {Kumar}, {Kuo}, {Kutynia}, {Kwee}, {Lackey}, {Landry}, {Lange},
  {Lantz}, {Lasky}, {Lazzarini}, {Lazzaro}, {Leaci}, {Leavey}, {Lebigot},
  {Lee}, {Lee}, {Lee}, {Lee}, {Lenon}, {Leonardi}, {Leong}, {Leroy},
  {Letendre}, {Levin}, {Levine}, {Li}, {Libson}, {Littenberg}, {Lockerbie},
  {Logue}, {Lombardi}, {London}, {Lord}, {Lorenzini}, {Loriette}, {Lormand},
  {Losurdo}, {Lough}, {Lousto}, {Lovelace}, {L{\"u}ck}, {Lundgren}, {Luo},
  {Lynch}, {Ma}, {MacDonald}, {Machenschalk}, {MacInnis}, {Macleod},
  {Maga{\~n}a-Sandoval}, {Magee}, {Mageswaran}, {Majorana}, {Maksimovic},
  {Malvezzi}, {Man}, {Mandel}, {Mandic}, {Mangano}, {Mansell}, {Manske},
  {Mantovani}, {Marchesoni}, {Marion}, {M{\'a}rka}, {M{\'a}rka}, {Markosyan},
  {Maros}, {Martelli}, {Martellini}, {Martin}, {Martin}, {Martynov}, {Marx},
  {Mason}, {Masserot}, {Massinger}, {Masso-Reid}, {Matichard}, {Matone},
  {Mavalvala}, {Mazumder}, {Mazzolo}, {McCarthy}, {McClelland}, {McCormick},
  {McGuire}, {McIntyre}, {McIver}, {McManus}, {McWilliams}, {Meacher},
  {Meadors}, {Meidam}, {Melatos}, {Mendell}, {Mendoza-Gandara}, {Mercer},
  {Merilh}, {Merzougui}, {Meshkov}, {Messenger}, {Messick}, {Meyers},
  {Mezzani}, {Miao}, {Michel}, {Middleton}, {Mikhailov}, {Milano}, {Miller},
  {Millhouse}, {Minenkov}, {Ming}, {Mirshekari}, {Mishra}, {Mitra},
  {Mitrofanov}, {Mitselmakher}, {Mittleman}, {Moggi}, {Mohan}, {Mohapatra},
  {Montani}, {Moore}, {Moore}, {Moraru}, {Moreno}, {Morriss}, {Mossavi},
  {Mours}, {Mow-Lowry}, {Mueller}, {Mueller}, {Muir}, {Mukherjee}, {Mukherjee},
  {Mukherjee}, {Mukund}, {Mullavey}, {Munch}, {Murphy}, {Murray}, {Mytidis},
  {Nardecchia}, {Naticchioni}, {Nayak}, {Necula}, {Nedkova}, {Nelemans},
  {Neri}, {Neunzert}, {Newton}, {Nguyen}, {Nielsen}, {Nissanke}, {Nitz},
  {Nocera}, {Nolting}, {Normandin}, {Nuttall}, {Oberling}, {Ochsner}, {O'Dell},
  {Oelker}, {Ogin}, {Oh}, {Oh}, {Ohme}, {Oliver}, {Oppermann}, {Oram},
  {O'Reilly}, {O'Shaughnessy}, {Ott}, {Ottaway}, {Ottens}, {Overmier}, {Owen},
  {Pai}, {Pai}, {Palamos}, {Palashov}, {Palomba}, {Pal-Singh}, {Pan}, {Pan},
  {Pankow}, {Pannarale}, {Pant}, {Paoletti}, {Paoli}, {Papa}, {Paris},
  {Parker}, {Pascucci}, {Pasqualetti}, {Passaquieti}, {Passuello},
  {Patricelli}, {Patrick}, {Pearlstone}, {Pedraza}, {Pedurand}, {Pekowsky},
  {Pele}, {Penn}, {Perreca}, {Pfeiffer}, {Phelps}, {Piccinni}, {Pichot},
  {Pickenpack}, {Piergiovanni}, {Pierro}, {Pillant}, {Pinard}, {Pinto},
  {Pitkin}, {Poeld}, {Poggiani}, {Popolizio}, {Post}, {Powell}, {Prasad},
  {Predoi}, {Premachandra}, {Prestegard}, {Price}, {Prijatelj}, {Principe},
  {Privitera}, {Prix}, {Prodi}, {Prokhorov}, {Puncken}, {Punturo}, {Puppo},
  {P{\"u}rrer}, {Qi}, {Qin}, {Quetschke}, {Quintero}, {Quitzow-James}, {Raab},
  {Rabeling}, {Radkins}, {Raffai}, {Raja}, {Rakhmanov}, {Ramet}, {Rapagnani},
  {Raymond}, {Razzano}, {Re}, {Read}, {Reed}, {Regimbau}, {Rei}, {Reid},
  {Reitze}, {Rew}, {Reyes}, {Ricci}, {Riles}, {Robertson}, {Robie}, {Robinet},
  {Rocchi}, {Rolland}, {Rollins}, {Roma}, {Romano}, {Romano}, {Romanov},
  {Romie}, {Rosi{\'n}ska}, {Rowan}, {R{\"u}diger}, {Ruggi}, {Ryan}, {Sachdev},
  {Sadecki}, {Sadeghian}, {Salconi}, {Saleem}, {Salemi}, {Samajdar}, {Sammut},
  {Sampson}, {Sanchez}, {Sandberg}, {Sandeen}, {Sanders}, {Sanders},
  {Sassolas}, {Sathyaprakash}, {Saulson}, {Sauter}, {Savage}, {Sawadsky},
  {Schale}, {Schilling}, {Schmidt}, {Schmidt}, {Schnabel}, {Schofield},
  {Sch{\"o}nbeck}, {Schreiber}, {Schuette}, {Schutz}, {Scott}, {Scott},
  {Sellers}, {Sengupta}, {Sentenac}, {Sequino}, {Sergeev}, {Serna},
  {Setyawati}, {Sevigny}, {Shaddock}, {Shaffer}, {Shah}, {Shahriar}, {Shaltev},
  {Shao}, {Shapiro}, {Shawhan}, {Sheperd}, {Shoemaker}, {Shoemaker}, {Siellez},
  {Siemens}, {Sigg}, {Silva}, {Simakov}, {Singer}, {Singer}, {Singh}, {Singh},
  {Singhal}, {Sintes}, {Slagmolen}, {Smith}, {Smith}, {Smith}, {Smith}, {Son},
  {Sorazu}, {Sorrentino}, {Souradeep}, {Srivastava}, {Staley}, {Steinke},
  {Steinlechner}, {Steinlechner}, {Steinmeyer}, {Stephens}, {Stevenson},
  {Stone}, {Strain}, {Straniero}, {Stratta}, {Strauss}, {Strigin}, {Sturani},
  {Stuver}, {Summerscales}, {Sun}, {Sutton}, {Swinkels}, {Szczepa{\'n}czyk},
  {Tacca}, {Talukder}, {Tanner}, {T{\'a}pai}, {Tarabrin}, {Taracchini},
  {Taylor}, {Theeg}, {Thirugnanasambandam}, {Thomas}, {Thomas}, {Thomas},
  {Thorne}, {Thorne}, {Thrane}, {Tiwari}, {Tiwari}, {Tokmakov}, {Tomlinson},
  {Tonelli}, {Torres}, {Torrie}, {T{\"o}yr{\"a}}, {Travasso}, {Traylor},
  {Trifir{\`o}}, {Tringali}, {Trozzo}, {Tse}, {Turconi}, {Tuyenbayev},
  {Ugolini}, {Unnikrishnan}, {Urban}, {Usman}, {Vahlbruch}, {Vajente},
  {Valdes}, {Vallisneri}, {van Bakel}, {van Beuzekom}, {van den Brand}, {Van
  Den Broeck}, {Vander-Hyde}, {van der Schaaf}, {van Heijningen}, {van Veggel},
  {Vardaro}, {Vass}, {Vas{\'u}th}, {Vaulin}, {Vecchio}, {Vedovato}, {Veitch},
  {Veitch}, {Venkateswara}, {Verkindt}, {Vetrano}, {Vicer{\'e}}, {Vinciguerra},
  {Vine}, {Vinet}, {Vitale}, {Vo}, {Vocca}, {Vorvick}, {Voss}, {Vousden},
  {Vyatchanin}, {Wade}, {Wade}, {Wade}, {Waldman}, {Walker}, {Wallace},
  {Walsh}, {Wang}, {Wang}, {Wang}, {Wang}, {Wang}, {Ward}, {Ward}, {Warner},
  {Was}, {Weaver}, {Wei}, {Weinert}, {Weinstein}, {Weiss}, {Welborn}, {Wen},
  {We{\ss}els}, {Westphal}, {Wette}, {Whelan}, {Whitcomb}, {White}, {Whiting},
  {Wiesner}, {Wilkinson}, {Willems}, {Williams}, {Williams}, {Williamson},
  {Willis}, {Willke}, {Wimmer}, {Winkelmann}, {Winkler}, {Wipf}, {Wiseman},
  {Wittel}, {Woan}, {Worden}, {Wright}, {Wu}, {Yablon}, {Yakushin}, {Yam},
  {Yamamoto}, {Yancey}, {Yap}, {Yu}, {Yvert}, {Zadro{\.Z}ny}, {Zangrando},
  {Zanolin}, {Zendri}, {Zevin}, {Zhang}, {Zhang}, {Zhang}, {Zhang}, {Zhao},
  {Zhou}, {Zhou}, {Zhu}, {Zucker}, {Zuraw}, {Zweizig}, {LIGO Scientific
  Collaboration}, \& {Virgo Collaboration}}]{2016PhRvL.116f1102A}
{Abbott}, B.~P., {Abbott}, R., {Abbott}, T.~D., {et~al.} 2016, \prl, 116,
  061102, \dodoi{10.1103/PhysRevLett.116.061102}

\bibitem[{Artymowicz \& Lubow(1994)}]{artymowicz_dynamics_1994}
Artymowicz, P., \& Lubow, S.~H. 1994, The Astrophysical Journal, 421, 651,
  \dodoi{10.1086/173679}

\bibitem[{{Begelman} {et~al.}(1980){Begelman}, {Blandford}, \&
  {Rees}}]{1980Natur.287..307B}
{Begelman}, M.~C., {Blandford}, R.~D., \& {Rees}, M.~J. 1980, \nat, 287, 307,
  \dodoi{10.1038/287307a0}

\bibitem[{{Bowen} {et~al.}(2018){Bowen}, {Mewes}, {Campanelli}, {Noble},
  {Krolik}, \& {Zilh{\~a}o}}]{2018ApJ...853L..17B}
{Bowen}, D.~B., {Mewes}, V., {Campanelli}, M., {et~al.} 2018, \apjl, 853, L17,
  \dodoi{10.3847/2041-8213/aaa756}

\bibitem[{Copperwheat {et~al.}(2005)Copperwheat, Cropper, Soria, \&
  Wu}]{copperwheat_optical_2005}
Copperwheat, C., Cropper, M., Soria, R., \& Wu, K. 2005, Monthly Notices of the
  Royal Astronomical Society, 362, 79, \dodoi{10.1111/j.1365-2966.2005.09223.x}

\bibitem[{Copperwheat {et~al.}(2007)Copperwheat, Cropper, Soria, \&
  Wu}]{copperwheat_irradiation_2007}
---. 2007, Monthly Notices of the Royal Astronomical Society, 376, 1407,
  \dodoi{10.1111/j.1365-2966.2007.11551.x}

\bibitem[{{Cuadra} {et~al.}(2009){Cuadra}, {Armitage}, {Alexander}, \&
  {Begelman}}]{2009MNRAS.393.1423C}
{Cuadra}, J., {Armitage}, P.~J., {Alexander}, R.~D., \& {Begelman}, M.~C. 2009,
  \mnras, 393, 1423, \dodoi{10.1111/j.1365-2966.2008.14147.x}

\bibitem[{{DeLaurentiis} {et~al.}(2024){DeLaurentiis}, {Haiman},
  {Westernacher-Schneider}, {Major Krauth}, {Davelaar}, {Zrake}, \&
  {MacFadyen}}]{2024arXiv240507897D}
{DeLaurentiis}, S., {Haiman}, Z., {Westernacher-Schneider}, J.~R., {et~al.}
  2024, arXiv e-prints, arXiv:2405.07897, \dodoi{10.48550/arXiv.2405.07897}

\bibitem[{{Dittmann} {et~al.}(2023){Dittmann}, {Ryan}, \&
  {Miller}}]{2023ApJ...949L..30D}
{Dittmann}, A.~J., {Ryan}, G., \& {Miller}, M.~C. 2023, \apjl, 949, L30,
  \dodoi{10.3847/2041-8213/acd183}

\bibitem[{D'Orazio {et~al.}(2016)D'Orazio, Haiman, Duffell, MacFadyen, \&
  Farris}]{dorazio_transition_2016}
D'Orazio, D.~J., Haiman, Z., Duffell, P., MacFadyen, A., \& Farris, B. 2016,
  Monthly Notices of the Royal Astronomical Society, 459, 2379,
  \dodoi{10.1093/mnras/stw792}

\bibitem[{{D'Orazio} {et~al.}(2013){D'Orazio}, {Haiman}, \&
  {MacFadyen}}]{2013MNRAS.436.2997D}
{D'Orazio}, D.~J., {Haiman}, Z., \& {MacFadyen}, A. 2013, \mnras, 436, 2997,
  \dodoi{10.1093/mnras/stt1787}

\bibitem[{{D'Orazio} {et~al.}(2015){D'Orazio}, {Haiman}, \&
  {Schiminovich}}]{2015Natur.525..351D}
{D'Orazio}, D.~J., {Haiman}, Z., \& {Schiminovich}, D. 2015, \nat, 525, 351,
  \dodoi{10.1038/nature15262}

\bibitem[{{Eggleton}(1983)}]{1983ApJ...268..368E}
{Eggleton}, P.~P. 1983, \apj, 268, 368, \dodoi{10.1086/160960}

\bibitem[{{Farris} {et~al.}(2014){Farris}, {Duffell}, {MacFadyen}, \&
  {Haiman}}]{2014ApJ...783..134F}
{Farris}, B.~D., {Duffell}, P., {MacFadyen}, A.~I., \& {Haiman}, Z. 2014, \apj,
  783, 134, \dodoi{10.1088/0004-637X/783/2/134}

\bibitem[{{Fukue}(1992)}]{1992PASJ...44..663F}
{Fukue}, J. 1992, \pasj, 44, 663

\bibitem[{{Haiman} {et~al.}(2009){Haiman}, {Kocsis}, \&
  {Menou}}]{2009ApJ...700.1952H}
{Haiman}, Z., {Kocsis}, B., \& {Menou}, K. 2009, \apj, 700, 1952,
  \dodoi{10.1088/0004-637X/700/2/1952}

\bibitem[{{Hayakawa}(1981)}]{1981PASJ...33..365H}
{Hayakawa}, S. 1981, \pasj, 33, 365

\bibitem[{{Hayasaki}(2009)}]{2009PASJ...61...65H}
{Hayasaki}, K. 2009, \pasj, 61, 65, \dodoi{10.1093/pasj/61.1.65}

\bibitem[{Hayasaki {et~al.}(2008)Hayasaki, Mineshige, \&
  Ho}]{hayasaki_supermassive_2008}
Hayasaki, K., Mineshige, S., \& Ho, L.~C. 2008, The Astrophysical Journal, 682,
  1134, \dodoi{10.1086/588837}

\bibitem[{Hayasaki {et~al.}(2007)Hayasaki, Mineshige, \&
  Sudou}]{hayasaki_binary_2007}
Hayasaki, K., Mineshige, S., \& Sudou, H. 2007, Publications of the
  Astronomical Society of Japan, 59, 427, \dodoi{10.1093/pasj/59.2.427}

\bibitem[{{Hayasaki} {et~al.}(2016){Hayasaki}, {Takahashi}, {Sendouda}, \&
  {Nagataki}}]{2016PASJ...68...66H}
{Hayasaki}, K., {Takahashi}, K., {Sendouda}, Y., \& {Nagataki}, S. 2016, \pasj,
  68, 66, \dodoi{10.1093/pasj/psw065}

\bibitem[{Hayasaki {et~al.}(2013)Hayasaki, Yagi, Tanaka, \&
  Mineshige}]{hayasaki_gravitational_2013}
Hayasaki, K., Yagi, K., Tanaka, T., \& Mineshige, S. 2013, Physical Review D,
  87, 044051, \dodoi{10.1103/PhysRevD.87.044051}

\bibitem[{{Ioka} {et~al.}(1998){Ioka}, {Chiba}, {Tanaka}, \&
  {Nakamura}}]{1998PhRvD..58f3003I}
{Ioka}, K., {Chiba}, T., {Tanaka}, T., \& {Nakamura}, T. 1998, \prd, 58,
  063003, \dodoi{10.1103/PhysRevD.58.063003}

\bibitem[{Kato {et~al.}(1998)Kato, Fukue, \& Mineshige}]{kato_black-hole_1998}
Kato, S., Fukue, J., \& Mineshige, S. 1998, Black-hole accretion disks.
\newblock \url{https://ui.adsabs.harvard.edu/abs/1998bhad.conf.....K}

\bibitem[{Kato {et~al.}(2008)Kato, Fukue, \& Mineshige}]{kato_black-hole_2008}
---. 2008, Black-{Hole} {Accretion} {Disks} --- {Towards} a {New} {Paradigm}
  ---.
\newblock \url{https://ui.adsabs.harvard.edu/abs/2008bhad.book.....K}

\bibitem[{{King} \& {Ritter}(1998)}]{1998MNRAS.293L..42K}
{King}, A.~R., \& {Ritter}, H. 1998, \mnras, 293, L42,
  \dodoi{10.1046/j.1365-8711.1998.01295.x}

\bibitem[{{MacFadyen} \& {Milosavljevi{\'c}}(2008)}]{2008ApJ...672...83M}
{MacFadyen}, A.~I., \& {Milosavljevi{\'c}}, M. 2008, \apj, 672, 83,
  \dodoi{10.1086/523869}

\bibitem[{Matsumoto \& Fukue(1998)}]{matsumoto_irradiated_1998}
Matsumoto, K., \& Fukue, J. 1998, Publications of the Astronomical Society of
  Japan, 50, 89, \dodoi{10.1093/pasj/50.1.89}

\bibitem[{{Miller} \& {Hamilton}(2002)}]{2002ApJ...576..894M}
{Miller}, M.~C., \& {Hamilton}, D.~P. 2002, \apj, 576, 894,
  \dodoi{10.1086/341788}

\bibitem[{{Noble} {et~al.}(2012){Noble}, {Mundim}, {Nakano}, {Krolik},
  {Campanelli}, {Zlochower}, \& {Yunes}}]{2012ApJ...755...51N}
{Noble}, S.~C., {Mundim}, B.~C., {Nakano}, H., {et~al.} 2012, \apj, 755, 51,
  \dodoi{10.1088/0004-637X/755/1/51}

\bibitem[{{Peters}(1964)}]{Peters64}
{Peters}, P.~C. 1964, Physical Review, 136, 1224,
  \dodoi{10.1103/PhysRev.136.B1224}

\bibitem[{{Pringle}(1981)}]{1981ARA&A..19..137P}
{Pringle}, J.~E. 1981, \araa, 19, 137,
  \dodoi{10.1146/annurev.aa.19.090181.001033}

\bibitem[{Rafikov(2013)}]{rafikov_structure_2013}
Rafikov, R.~R. 2013, The Astrophysical Journal, 774, 144,
  \dodoi{10.1088/0004-637X/774/2/144}

\bibitem[{Rafikov(2016)}]{rafikov_generalized_2016}
---. 2016, The Astrophysical Journal, 830, 7, \dodoi{10.3847/0004-637X/830/1/7}

\bibitem[{Rahoui {et~al.}(2010)Rahoui, Chaty, Rodriguez, Fuchs, Mirabel, \&
  Pooley}]{rahoui_long-term_2010}
Rahoui, F., Chaty, S., Rodriguez, J., {et~al.} 2010, The Astrophysical Journal,
  715, 1191, \dodoi{10.1088/0004-637X/715/2/1191}

\bibitem[{{Roedig} {et~al.}(2014){Roedig}, {Krolik}, \&
  {Miller}}]{2014ApJ...785..115R}
{Roedig}, C., {Krolik}, J.~H., \& {Miller}, M.~C. 2014, \apj, 785, 115,
  \dodoi{10.1088/0004-637X/785/2/115}

\bibitem[{{Sanbuichi} {et~al.}(1993){Sanbuichi}, {Yamada}, \&
  {Fukue}}]{1993PASJ...45..443S}
{Sanbuichi}, K., {Yamada}, T.~T., \& {Fukue}, J. 1993, \pasj, 45, 443

\bibitem[{{Sasaki} {et~al.}(2016){Sasaki}, {Suyama}, {Tanaka}, \&
  {Yokoyama}}]{2016PhRvL.117f1101S}
{Sasaki}, M., {Suyama}, T., {Tanaka}, T., \& {Yokoyama}, S. 2016, \prl, 117,
  061101, \dodoi{10.1103/PhysRevLett.117.061101}

\bibitem[{{Shakura} \& {Sunyaev}(1973)}]{1973A&A....24..337S}
{Shakura}, N.~I., \& {Sunyaev}, R.~A. 1973, \aap, 24, 337

\bibitem[{{Shi} \& {Krolik}(2015)}]{2015ApJ...807..131S}
{Shi}, J.-M., \& {Krolik}, J.~H. 2015, \apj, 807, 131,
  \dodoi{10.1088/0004-637X/807/2/131}

\bibitem[{{Sutton} {et~al.}(2014){Sutton}, {Done}, \&
  {Roberts}}]{2014MNRAS.444.2415S}
{Sutton}, A.~D., {Done}, C., \& {Roberts}, T.~P. 2014, \mnras, 444, 2415,
  \dodoi{10.1093/mnras/stu1597}

\end{thebibliography}
\bibliographystyle{aasjournal}
%

\end{document}